\documentclass[11pt,a4paper]{article} 
\pdfoutput=1
\usepackage{style}

\usepackage{amsmath}
\usepackage{amssymb}
\usepackage{amsthm}
\usepackage{psfrag}
\usepackage{graphicx}
\usepackage{hyperref}
\usepackage{feynmp}
\usepackage{color}
\usepackage{bm}

\newcommand{\be}{\begin{equation}}
\newcommand{\ee}{\end{equation}}
\newcommand{\beqa}{\begin{eqnarray}}
\newcommand{\eeqa}{\end{eqnarray}}

\newcommand\Wn{\Gamma^{tot}}
\newcommand{\W}{{\Gamma}}
\newcommand\G{\Gamma}

\newcommand\Ct{C'}

\newcommand\s{\sigma}

\renewcommand\a{\alpha}
\renewcommand\b{\beta}

\newcommand{\T}{\Theta}

\newcommand{\HH}{{\cal H}}

\newcommand\x{{\bf x}}
\renewcommand\k{{\bf k}}
\newcommand\q{{\bf q}}
\newcommand{\e}{\eta}

\def\d{\partial}
\newcommand{\bseq}{\begin{subequations}}
\newcommand{\eseq}{\end{subequations}}

\renewcommand{\ln}{\mathop{\rm ln}\nolimits}

\relax

\title{Time-Sliced Perturbation Theory 
for Large Scale Structure I: General Formalism}

\author[a,1]{Diego Blas,\!\note{diego.blas@cern.ch}}
\author[a,2]{Mathias Garny,\!\note{mathias.garny@cern.ch}}
\author[b,c,d,3]{Mikhail M. Ivanov\note{mikhail.ivanov@cern.ch}}
\author[a,b,d,4]{and Sergey Sibiryakov,\!\note{sergey.sibiryakov@cern.ch}}

\affiliation[a]{CERN Theory Division, CH-1211 Gen\`eve 23, Switzerland}
\affiliation[b]{FSB/ITP/LPPC, \'Ecole Polytechnique F\'ed\'erale de Lausanne, \normalsize\it CH-1015, Lausanne, Switzerland}
\affiliation[c]{Faculty of Physics, Moscow State University,\normalsize \it Vorobjevy Gory, 119991 Moscow, Russia}
\affiliation[d]{Institute for Nuclear Research of the
Russian Academy of Sciences, \\ 
\normalsize \it  60th October Anniversary Prospect, 7a, 117312
Moscow, Russia}

\abstract{We present a new analytic approach to describe large scale
  structure formation in the mildly non-linear regime. The central
  object of the method is the time-dependent probability distribution
  function generating correlators of the cosmological observables at a
  given moment of time. Expanding the distribution function around the
  Gaussian weight we formulate a perturbative technique to calculate
  non-linear corrections to cosmological correlators, similar to the
  diagrammatic expansion in a three-dimensional Euclidean quantum
  field theory, with time playing the role of an external
  parameter. For the physically relevant case of cold dark matter in an
  Einstein--de Sitter universe, the time evolution of the distribution
  function can be found exactly and is encapsulated by a
  time-dependent coupling constant  controlling the perturbative
  expansion.  
We show that all building blocks of the expansion are free from spurious
infrared enhanced contributions that plague the standard cosmological
perturbation theory. This paves the way towards the systematic resummation
of infrared effects in large scale structure formation. We also argue
that the approach proposed here provides a natural framework to account for the
influence of short-scale dynamics on larger scales along the lines of
effective field theory.} 

\begin{document}

\begin{flushright}
CERN-PH-TH-2015-298\\
INR-TH-2015-034
\end{flushright}

\maketitle


\section{Introduction}

Ongoing and future cosmological surveys will provide an unprecedented amount of data about the structure of the universe at large scales. The statistical properties of this large scale structure (LSS) are believed to
contain a wealth of information about the 
primordial constituents and dynamical evolution of the universe. 
However, harvesting this information will not be an easy task and will require, in particular, a precise theoretical understanding of self-gravitating systems far away from equilibrium.  
The most straightforward approach to this problem relies on current computer power to produce realistic $N$-body simulations for a large range of scales. 
Impressive as they are, these simulations are still very time-consuming, which calls for (semi-)analytic methods.  At the largest scales, above {\em a few} megaparsecs, the dynamics is dominated by dark matter which, in turn, can be described as an almost perfect pressureless fluid, possibly with stochastic sources \cite{Peebles:1980,Pueblas:2008uv,Baumann:2010tm,Pietroni:2011iz,Carrasco:2012cv,Baldauf:2015tla}.   
At these scales the matter density contrast is a small quantity which justifies the use of perturbative techniques. 

Standard perturbation theory (SPT) of LSS formation \cite{Bernardeau:2001qr,bernardeau} is one of the most popular of these techniques. It consists of two main steps: first, one expresses the dark matter
density- and velocity fields at a given time as a power series of the
initial conditions, assuming a perfect pressureless fluid without vorticity. Next, one 
performs the ensemble averages using the statistical distribution at the
initial time when the system is well within the linear regime. The initial distribution is often taken to be Gaussian\footnote{It is worth stressing that the precise form of the distribution should be provided by the theory describing the generation of primordial fluctuations. Disentangling the primordial non-Gaussianity from the secondary one induced by non-linear dynamics constitutes one of the goals of the LSS studies.}, as motivated by the constraints coming  from the cosmic microwave background measurements \cite{Ade:2015ava}. This framework leads to a loop expansion for non-linear corrections to cosmological correlation functions and has provided numerous insights into their properties. 

However, SPT possesses a number of drawbacks which have recently
attracted significant attention. They can be attributed to the sensitivity of the SPT computational scheme to
the infrared (IR) and ultraviolet (UV) modes. The presence of mode-mixing is  a consequence of the  non-linear dynamics. 
In the loop integrals of SPT this effect receives large contributions both from very small (IR) and very large (UV) 
wavenumbers. 
Technically, the individual loop integrals in SPT would be IR (UV) divergent for an initial power spectrum\footnote{We refer as `initial' to quantities defined after recombination, that serve as the input for non-linear structure formation. } $P(k)$ behaving as $k^\nu$ with 
$\nu\leq -1$ ($\nu>-3$) at $k\to 0$ ($k\to\infty$)\footnote{More precisely, UV divergences arise at $L$-loop
order if $\nu\geq -3+2/L$ for $k\to\infty$ \cite{Goroff:1986ep}.}. 
Realistic $\Lambda$CDM spectra avoid these conditions and divergences are absent. 
However,
the corresponding
IR/UV regions in the loop integrals still give rise to large contributions 
 that ultimately limit the applicability of perturbative computations. They are somewhat loosely 
referred to as ``IR/UV divergences" even for $\Lambda$CDM.

Qualitatively the appearance of IR divergences in SPT stems from the use of the initial distribution to evaluate quantities at late times. 
This introduces non-local time dependence 
to the large displacements of the fluid particles caused by large scale bulk flows.
It is well known that the IR divergences cancel out
in equal-time correlators upon summing over all sub-diagrams at a fixed order in perturbation theory \cite{Vishniac}. This cancellation has been formally proven for \emph{leading} IR divergences to all orders of perturbation theory  \cite{Jain:1995kx} and can be traced back to
the equivalence principle \cite{Scoccimarro:1995if,Creminelli:2013mca}. 
Recently, the cancellation has been proven also for \emph{subleading} IR divergences showing up for the first time at 2 loops \cite{Blas:2013bpa, Carrasco:2013sva, Sugiyama:2013gza, Kehagias:2013yd, Peloso:2013zw}.
Still, the presence of spurious IR divergences greatly complicates numerical calculations\footnote{IR-safe integrands have been constructed in \cite{Blas:2013bpa, Carrasco:2013sva} for 2 loops and in \cite{Blas:2013aba} for an arbitrary $L$-loop order.} and 
obscures the analysis of physical effects produced by the large scale bulk flows. The latter, though finite, have a strong impact on the features in the cosmological correlation functions \cite{Eisenstein:2006nj}
In particular, a resummation of physical IR contributions is essential 
 for an accurate description of baryon acoustic oscillations (BAO) in the power spectrum \cite{Eisenstein:2006nj,Crocce:2007dt,Senatore:2014via, Baldauf:2015xfa}.

Another problem is related to the UV sensitivity of the SPT loop
expansion \cite{bernardeau,Blas:2013aba}. 
This makes the perturbative results at high loop order 
largely dependent on the short wavelength modes. In particular, the calculations are very sensitive to the scales
where  the fluid approach fails. This issue was recently addressed by applying ideas of effective theories to LSS
  \cite{Carrasco:2012cv}. In these approaches, one 
renormalizes the UV contributions and parameterizes the ignorance
about the dynamics at short scales by various effective
operators in the equations of
motion of the dark matter fluid which are 
fixed from data or $N$-body simulations, see  \cite{Baldauf:2015aha,Foreman:2015lca}. They exhibit non-local
time dependence, 
which complicates the renormalisation at high loop order  \cite{Abolhasani:2015mra}.
Yet, the results from  $N$-body simulations show that the
actual sensitivity of the power spectrum to UV modes is smaller than
what is observed in SPT \cite{Pueblas:2008uv,Nishimichi:2014rra, Garny:2015oya}, 
confirming the expectations based on qualitative 
arguments \cite{Peebles:1980}. This suggests that an accurate
description of LSS based on perturbation theory may be possible under rather broad assumptions
about the short-scale dynamics.
Therefore, one is motivated to develop approaches where the UV contributions
can be consistently isolated and their sensitivity to various assumptions
systematically studied \cite{Pietroni:2011iz,Blas:2015tla,Fuhrer:2015cia}.

In this paper, we advocate a new framework within Eulerian
hydrodynamics 
which can overcome the aforementioned drawbacks of SPT.
The main idea is to evolve the statistical
distribution function of the fields rather than the fields themselves.
This is done using the Liouville equation of statistical mechanics. The
perturbation theory is then developed over the Gaussian part of the
distribution at final time.
The latter closely resembles perturbation theory in a (non-local)
3-dimensional Euclidean quantum field theory (QFT), 
with time playing the role of an external parameter\footnote{We will see
  that for Einstein--de Sitter cosmology  time enters
through a time-dependent coupling constant controlling the
perturbative expansion. This is also a very
  good approximation in the   $\Lambda$CDM case.}.  
In this way, we disentangle time evolution from statistical
averaging. We denote the formalism by 
\textit{Time-Sliced Perturbation Theory} (TSPT). 
We are going to argue that it is well adapted for calculating
cosmological correlators with all entries taken at the same
time. These observables are of particular interest in cosmological observations \cite{Bernardeau:2001qr}.

The strategy to evolve the distribution function can in principle be
applied to any underlying evolution equations, 
and we pay special attention to keep the derivation general. 
For the equations of a pressureless
perfect fluid and at a fixed order of perturbation theory the TSPT
approach gives the 
same results as SPT. 
Nonetheless, we will show that TSPT has the important advantage 
that all building blocks of the diagrammatic expansion,
i.e. propagators and vertices, as well as individual diagrams
themselves are free from IR divergences. 
This can be traced back to the property that within TSPT one deals
only with equal-time objects which are protected from spurious divergences by the equivalence
principle. 
 TSPT is thus  a convenient
framework for implementing IR resummation --- a subject that 
is addressed in \cite{BAOnext}. On the UV side, TSPT allows to 
reformulate the effective field
theory of LSS in the language of Wilsonian renormalization group
within the 3-dimensional Euclidean QFT describing the statistical
averaging.  This formulation
appears promising to shed new light on the
properties of the effective operators.
The inclusion of stochastic noise in the evolution
\cite{Carrasco:2012cv,Baldauf:2015tla} can, in principle, be
incorporated by promoting 
the Liouville equation for the distribution function to a Fokker-Planck equation.
Finally, the structure of TSPT is also 
suitable to include primordial
non-Gaussianity in a straightforward 
manner.

The paper is organized as follows. In sec.~\ref{sec:fluid} we recall
the 
Eulerian fluid equations, define the Zel'dovich approximation and fix
our notations. The general formalism is introduced in  
sec.~\ref{sec:tspt}, where we present the strategy to solve the
Liouville equation and derive the perturbative expansion for
correlation functions. We apply this method 
to the Zel'dovich approximation in 
sec.~\ref{sec:zel} and to the exact Eulerian dynamics in sec.~\ref{sec:ED}.  
In sec.~\ref{sec:IRsafe} we prove the IR finiteness of the vertex
functions and relate it to the equivalence principle. 
Section \ref{sec:disc} is devoted to discussion and
future directions. Formulas relating the TSPT vertex functions
with the SPT 
kernels are given in 
appendix~\ref{app:spttotspt}. Appendix~\ref{app:1loop} contains 
an explicit example of a one-loop computation in TSPT and its comparison
to the SPT result. 
Finally, appendix~\ref{app:initialPS} is devoted to some technical 
details regarding the IR - safety of TSPT vertices.

\section{Perfect fluid evolution equations}\label{sec:fluid}

In this section we fix our notations and conventions by briefly reminding
the basic equations underlying the SPT approach (see
\cite{Bernardeau:2001qr,bernardeau} for a detailed discussion).
In general, the evolution of non-relativistic, gravitationally interacting
matter in an expanding universe is described by the Vlasov--Poisson
equation for the
phase-space distribution 
function. By taking moments of the distribution function over the velocity one
obtains a coupled hierarchy 
of evolution equations. Its truncation leads to the Eulerian equations for a
perfect pressureless fluid, 
\bseq
\label{hydro}
\begin{align}
 & \frac{\d\delta}{\d t} + \nabla \cdot [(1+\delta){\bm u}] = 0 \,, \\
 & \frac{\d{\bm u}}{\d t} + {\cal H}{\bm u} + ({\bm u}\cdot \nabla) {\bm
   u} =  -\nabla\Phi \,,
\end{align}
where $t$ is the conformal time, $\delta(t,\x)$ is the overdensity
field 
and ${\bm u}(t,\x)$ is the peculiar flow velocity. The gravitational
potential is determined by the Poisson equation 
\be
\label{newton}
\nabla^2\Phi=\frac{3}{2}{\cal H}^2\Omega_m\delta\;,
\ee 
\eseq
where ${\cal H}=aH$ is the conformal
Hubble rate, and $\Omega_m$ is the time-dependent matter density
parameter.
It is convenient to introduce a new time variable 
\[
\e\equiv \log D_+(t)\;,
\] 
where $D_+(t)$ is the growth factor of the linearized perturbations, 
and rewrite eqs.~(\ref{hydro}) in Fourier space.
Neglecting vorticity, which decays at the linear level and is not
sourced by the gradient force, we obtain,
\begin{subequations}
\label{psidotED}
\begin{align}
\label{psidotED1}
&\dot\delta_\e(\k)-\T_\e(\k)=\int [dq]^2\delta^{(3)}(\k-\q_1-\q_2) \a(\q_1,\q_2)\T_\e(\q_1)\delta_\e(\q_2),\\
 &\dot \T_\e(\k) -\frac{3\Omega_m}{2f^2}\delta_\e(\k)
+\!\bigg(\frac{3\Omega_m}{2f^2}-1\bigg)\T_\e(\k)
\!=\!\!\int[ d
q]^2\delta^{(3)}(\k-\q_1-\q_2)\b(\q_1,\q_2)\T_\e(\q_1)\T_\e(\q_2)\,,
\label{psidotED2}
\end{align}
\end{subequations}
where dot stands for a derivative with respect to $\eta$, and we have
introduced the notations,
\be
\delta_\e(\k)\equiv \delta(\e,\k)~,~~~~
\T_\e(\k)\equiv -\frac{\nabla{\bm u}(\e,\k)}{f(\e)\,\HH(\e)}~~~~~~
\text{with} ~~~ f\equiv \frac{d \ln D_+(\eta)}{d \ln a (\eta)}\;.
\ee
Note that we have rescaled the velocity divergence to minimize the
$\e$-dependence in the equations.
The non-linear kernels are given by
\be
\label{alphabeta}
\alpha(\k_1,\k_2)\equiv\frac{(\k_1+\k_2)\cdot \k_1}{k_1^2}\,, \quad \quad 
\b(\k_1,\k_2)\equiv\frac{(\k_1+\k_2)^2(\k_1\cdot \k_2)}{2k_1^2k_2^2}\,. 
\ee
Finally, we have used a shorthand notation for the
integration measure, $[d q]^n\equiv d^3q_1 \cdots d^3q_n$. 

In a matter-dominated universe (Einstein--de Sitter background) one
has $\Omega_m=f=1$ and the coefficients in eqs.~(\ref{psidotED}) are
time-independent. For  $\Lambda$CDM cosmology
the ratio $\Omega_m/f^2$ is also close to one and we will
simplify eqs.~(\ref{psidotED}) by replacing $\Omega_m/f^2\mapsto
1$. As shown in \cite{Pietroni:2008jx} (see also \cite{Bernardeau:2001qr}), the  error in the power spectrum
introduced by this replacement is less than $1\%$ at zero redshift
($z=0$) and less than $0.1\%$ at $z=1$ for all relevant scales. Thus, 
this replacement provides an accurate approximation to
the exact dynamics and the deviation of $\Omega_m/f^2$
from $1$ can be taken into account perturbatively. With slight abuse of
language we refer to the case described by eqs.~(\ref{psidotED})
with $\Omega_m/f^2=1$  as ``exact dynamics'' (ED).

Below we also consider the Zel'dovich approximation (ZA) which is
obtained by replacing $\delta$ in (\ref{psidotED2}) with $\T$. In this
case (\ref{psidotED2}) becomes a closed equation for $\T$. While in ZA
the correlation functions of cosmological observables can be found exactly, for us it is relevant as a testing ground
of our perturbative technique. ZA also captures correctly the
effects of large bulk flows in ED and hence both approaches share a similar  IR behavior \cite{Scoccimarro:1995if}. However, one should keep in mind that in
all other regimes the properties of ZA and ED are essentially different.

\section{The TSPT framework}\label{sec:tspt}

\subsection{Preliminaries}

In this section we develop the TSPT formulation for general evolution
equations, irrespectively of particular approximations for
the fluid dynamics. 
We start by considering a single field variable, which, to avoid
proliferation of notations, we denote with the same letter $\T$ as the
velocity divergence. It satisfies the deterministic equation of motion, 
\be
\label{eq:eomp}
\dot \T_\e(\k)={\cal I}[\T_\e;\e,\k]\,,
\ee
where the r.h.s. is given as a Taylor series in the fields,
\be
\label{eq:ik}
{\cal I}[\T_\e;\e,\k]=\sum_{n=1}^\infty \frac{1}{n!} \int[dq]^n \,
I_n\big(\e;\q_1,...,\q_n\big)\;
\delta^{(3)}\bigg(\k-\sum_{i=1}^n \q_i\bigg)\;
\prod_{j=1}^n \T_\e(\q_j).
\ee
The $\delta$-function in this expression enforces the conservation of
momentum, which we assume to be satisfied by the system. We also assume
that the system respects parity 
\be
I_m(\e;\q_1,...,\q_m)=I_m(\e;-\q_1,...,-\q_m) \,.
\ee

The second key ingredient is the statistical distribution at the
initial time $\e_0$ chosen deep in the linear regime. 
For illustrative purposes, let us assume that it is Gaussian. This is
a very good approximation for LSS, consistent with the Planck results 
\cite{Ade:2015ava}.
We will discuss how to include an initial non-Gaussianity later on.
According to this assumption, the correlators of $\T_\e$ are
determined by the generating functional \cite{ValageasN,eff1},  
\be
\label{eq:zt}
Z[J;\e]=\mathcal{N}^{-1}\int [\mathcal{D}\T_{\e_0}] 
\exp\bigg\{-\frac{1}{2}\int
[dk]\frac{\T_{\e_0}(\k)\T_{\e_0}(-\k)}{P_{\e_0}(|\k|)}
+\int[dk]\T_{\e}(\k) J(-\k)\bigg\}\,,
\ee
where
\be
 \mathcal{N}=\int [\mathcal{D}\T_{\e_0}] 
\exp\bigg\{-\frac{1}{2}\int
[dk]\frac{\T_{\e_0}(\k)\T_{\e_0}(-\k)}{P_{\e_0}(|\k|)} \bigg\}\,,
\ee
is the normalization factor. Note that the integration here is
performed over the fields at the initial time, whereas the source $J$
couples to the final values of the field. We have explicitly
implemented the statistical homogeneity and isotropy by postulating that
in the Gaussian weight the momenta of the $\T$-fields sum to zero and
the power spectrum $P_{\eta_0}(|\k|)$ depends only on the absolute
value of the momentum. The equal-time correlation functions are
obtained by varying $Z[J;\e]$ with respect to the source and setting
$J=0$ afterwards,
\be
\label{eq:corr}
\langle\T_\e(\k_1)\ldots\T_\e(\k_n)\rangle=
\frac{\delta^n Z[J;\e]}{\delta J(-\k_1)\ldots\delta J(-\k_n)}\bigg|_{J=0}\;.
\ee
In particular, for the two-point function at the initial time this
formula yields,
\be
\label{eq:dzdjj}
\langle \T_{\e_0}(\k_1)\T_{\e_0}(\k_2) \rangle
=\frac{\delta^2 Z[J;\e_0]}{\delta J(-\k_1) \delta J(-\k_2)}\bigg|_{J=0}
=P_{\e_0}(|\k_1|)\delta^{(3)}(\k_1+\k_2) \, .
\ee

In SPT the evolution equations (\ref{eq:eomp}) are solved iteratively
and the final field $\T_\e(\k)$ is expressed as a Taylor series in the
powers of the initial configuration $\T_{\e_0}(k)$, see
eq.~(\ref{eq:spt}) in appendix~\ref{app:spttotspt}. We suggest an
alternative procedure.
If one is only interested in correlation functions
of fields at a particular time, it seems natural to use them as the
main elements of the analysis. To do this, we substitute the
integration variable in \eqref{eq:zt} 
by the fields at time $\eta$, which defines a time-dependent
distribution function ${\mathcal P}[\T_\e;\e]$, 
\be
\label{eq:ztfp}
Z[J;\e]=\int [\mathcal{D}\T_\e]\;{\mathcal P}[\T_\e;\e]\;
\exp\bigg\{\int [dk] \T_\e(\k) J(-\k)\bigg\}\,.
\ee
The equation that determines the time evolution of ${\mathcal P}$ is
nothing but the classical \emph{Li\-ou\-ville} equation where the value
of the field at a particular $\k$ is considered as a statistical
variable. For a system obeying \eqref{eq:eomp} it reads,
\be
\label{eq:fp}
\frac{\d}{\d\e}{\cal P}[\T_\e;\e]+\int [d k] 
\frac{\delta}{\delta \T_\eta(\k)}
\Big({\cal I}[\T_\e;\e,\k] \,{\cal P}[\T_\e;\e]\Big)
=0\;.
\ee
This equation can be understood as the continuity equation for the
probability density in functional space. It can be derived by
performing a substitution of integration variables in \eqref{eq:ztfp}
in terms of fields at time  
$\eta+\delta\eta$, and taking $\delta\eta\to 0$ while demanding the invariance of
 generating functional on the choice of integration variables.
In what follows, we derive a recursive chain of equations to solve
\eqref{eq:fp}.
When ${\cal P}$ is found, one can compute the different correlation 
functions using (\ref{eq:ztfp}). 
In this way we disentangle the time evolution from the statistical averaging.
From now on we will omit the subindex ``$\e$'' on the field $\T$
whenever it appears as the argument of the distribution function.

\subsection{Vertices}

We search for the solution of (\ref{eq:ztfp}) in the form,
\be
{\cal P}[\T;\e]={\mathcal N}^{-1} \exp\big\{\!-\W[\T;\e]\big\}\,.
\ee
In the spirit of perturbation theory, we expand the statistical weight
$\Gamma$ as a series in the powers of the field $\T$,
\be
\label{eq:Wn}
{\W}[\T;\e]=\sum_{n=1}^{\infty}\frac{1}{n!}\int [dk]^n\;
\Wn_n(\e;\k_1,...,\k_n)\; \prod^n_{j=1} \T(\k_j) .
\ee
Substituting this into \eqref{eq:fp}, using the expression
(\ref{eq:ik})   and requiring that the
coefficients in front of equal powers of $\T$ vanish we obtain,
\be
\label{eq:Wdot}
\begin{split}
\dot \Gamma_n^{tot}(\e;\k_1,...,&\k_n)
+\sum_{m=1}^{n}\frac{1}{m!(n-m)!}\\
&\times\sum_{\sigma}I_{m}\big(\e;\k_{\s(1)},...,\k_{\s(m)}\big)
\Wn_{n-m+1}\Big(\e;\sum_{l=1}^m
\k_{\s(l)},\k_{\s(m+1)},...,\k_{\s(n)}\Big) 
\\
&=~\delta^{(3)}\left(\sum_{i=1}^n \k_i\right)
\int [dp]\, I_{n+1}\big(\e;{\mathbf p},\k_1,...,\k_n\big), 
\end{split}
\ee
where in the second term on the l.h.s. the sum runs over all
permutations $\s$ of $n$ indices. Note that, by construction, $I_n$,
$\Gamma^{tot}_n$ are symmetric functions of momenta, and thus the
terms obtained by permutations inside the first $m$ or last $(n-m)$
momenta are the same. In other words, the number of distinct terms in
the sum over $\s$ is equal to the binomial coefficient 
$\binom{n}{m}$ instead of $n!$. 
The chain of equations (\ref{eq:Wdot}) can be compared to the
so-called BBGKY hierarchy for the correlation functions appearing in
the standard approach 
\cite{Peebles:1980}. A key difference is that in our case the $n$-th
equation contains only the functions $\Gamma^{tot}_m$
with $m\leq n$ and hence can be solved exactly. 
In contrast, in the BBGKY hierarchy the equation for the $n$-point
function involves higher correlators. To solve it, one has
to truncate 
the hierarchy at a finite order, thereby introducing an error in the
solution. Still, the approach can be useful to solve SPT perturbatively \cite{Audren:2011ne}.

We will see in sec.~\ref{Sec:applic} that for the evolution equations
describing the dynamics of LSS the momentum integrals appearing on the
r.h.s. of eq.~(\ref{eq:Wdot}) are UV divergent. Thus, it is convenient
to split $\Wn_n$ into a regular part $\Gamma_n$ and a singular
`counterterm' $C_n$,
\be
\Wn_n= \G_n+C_n.
\ee
The vertex functions $\G_n$ satisfy \eqref{eq:Wdot} without sources
 and are subject to initial conditions encoding the statistical
properties of the primordial fluctuations. The $C_n$ 
satisfy trivial initial conditions and are sourced by the divergent
r.h.s. in (\ref{eq:Wdot}). As we will discuss shortly, they cancel
certain UV divergent contributions in the diagrammatic expansion based
on (\ref{eq:ztfp}), hence the name `counterterms'. The divergences stem from
the (singular) Jacobian describing the change in the functional
measure when going from (\ref{eq:zt}) to (\ref{eq:ztfp}).

To proceed, let us assume that the evolution kernels $I_n$ are
time-independent. As discussed in sec.~\ref{sec:fluid}, this is a good
approximation in the case of LSS. We further take 
\be
\label{eq:I1}
I_1=1\;,
\ee
which implies that the solution of linearized evolution equations
(\ref{eq:eomp}) grows as ${\rm e}^\eta$ uniformly at all momenta. This
is the case for a perfect pressureless fluid.\footnote{A momentum
  dependence of the growth factor  appears beyond the perfect fluid
  approximation \cite{Baumann:2010tm,Blas:2015tla}.} 
Let us first focus on the regular vertices $\G_n$. It is easy to see that 
$\G_1$ can be consistently set to $0$ at all times, which corresponds
to vanishing one-point function $\langle\T_\e\rangle=0$. For $n\geq 2$
we use the Ansatz,
\be
\label{GamAns}
\G_n\big(\eta;\k_1,\ldots,\k_n\big)=\sum_{l=2}^n{\rm e}^{-l\e}\;
\G_n^{(l)}(\k_1,\ldots,\k_n)\;.
\ee
Substituting it into (\ref{eq:Wdot}) with vanishing r.h.s. yields a
chain of relations,
\be
\label{Gnl}
\begin{split}
\G_n^{(l)}(\k_1,\ldots,\k_n)&=-\frac{1}{n-l}\sum_{m=2}^{n-l+1}
\frac{1}{m!(n-m)!}\\
&\times\sum_\s
I_m\big(\k_{\s(1)},\ldots,\k_{\s(m)}\big)\,
\G^{(l)}_{n-m+1}
\Big(\sum_{i=1}^m\k_{\s(i)},\k_{\s(m+1)},\ldots,\k_{\s(n)}\Big)\,,
\end{split}
\ee
for $2\leq l< n$, whereas $\G_n^{(n)}$ is arbitrary. The latter
must be determined by the initial conditions on $\G_n$. To simplify
the formulas, it is convenient to send the initial time $\e_0$ to $-\infty$. We obtain,
\be
\label{indata}
\G_n^{(n)}(\k_1,\ldots,\k_n)=\lim_{\e_0\to-\infty}
{\rm e}^{n\e_0}\;\G_n(\eta_0;\k_1,\ldots,\k_n)\;.
\ee

The solution (\ref{GamAns}) is greatly simplified in the physically
relevant case of Gaussian initial conditions. According to
(\ref{indata}) the initial data in this case read,
\bseq
\label{initGaus}
\begin{align}
\label{initGaus1}
&\G_2^{(2)}(\k_1,\k_2)=\frac{\delta^{(3)}(\k_1+\k_2)}{\bar
  P(|\k_1|)}\;,\\
&\G_n^{(n)}=0~,~~~~n>2\;.
\label{initGaus2}
\end{align}
\eseq
Here 
\be
\label{PSbar}
\bar P(|\k|)\equiv\lim_{\e_0\to-\infty}{\rm e}^{-2\e_0}\;P_{\e_0}(|\k|)
\ee
is the suitably rescaled initial power spectrum. Next, from (\ref{Gnl}) one
infers that all $\G_n^{(l)}$ with $l>2$ vanish, which leads to the
solution
\bseq
\label{solGaus}
\begin{align}
\label{solGaus1}
\G_n\big(\eta;\k_1,\ldots,&\k_n\big)={\rm e}^{-2\e}\;
\bar\G_n(\k_1,\ldots,\k_n)\;,\\
\bar\G_n(\k_1,\ldots,&\k_n)=-\frac{1}{n-2}\sum_{m=2}^{n-1}
\frac{1}{m!(n-m)!}\notag\\
&\times\sum_\s
I_m\big(\k_{\s(1)},\ldots,\k_{\s(m)}\big)\,
\bar\G_{n-m+1}
\Big(\sum_{i=1}^m\k_{\s(i)},\k_{\s(m+1)},\ldots,\k_{\s(n)}\Big)\,,
\label{solGaus2}
\end{align}
\eseq
where we have introduced the notation $\bar
\G_n\equiv\G_{n}^{(2)}$. One observes that all vertices are
proportional to the same factor ${\rm e}^{-2\e}$. This implies that
the time dependence factors out of the regular part of 
statistical weight\footnote{The
time-dependence of the counterterms $C_n$ is different, see eq.~\eqref{counter}.}
(\ref{eq:Wn}). It is suggestive to write it in the form,
\be
\label{treeact}
\G^{reg}[\T;\e]=\frac{1}{g^2(\e)} \bar\G[\T]\;,
\ee
where
\be
\label{coupling}
g(\e)\equiv{\rm e}^\e
\ee
and $\bar\G[\T]$ is time-independent. This expression implies that $g$
plays the role of the \emph{coupling constant} controlling the perturbative
expansion of the generating functional (\ref{eq:ztfp}). Note that $g$
grows with time, so that the perturbation theory breaks down at late
times, as expected for the dynamics of gravitational clustering.
Practical computations up to a fixed order in perturbative expansion
require the knowledge of only a few lowest-order vertices. These can
be easily found from the recursion relations
(\ref{solGaus2}) with the seed two-point function $\bar\G_2$ given by
(\ref{initGaus1}). Explicit expressions for  the three- and
four-point vertices are given in appendix~\ref{app:1loop}.

It is clear how to include initial non-Gaussianity in this
framework. For instance,  the presence of an initial bispectrum gives rise
to non-vanishing vertex $\G_3^{(3)}$. Through eqs.~(\ref{Gnl}) this
will generate a sequence of descendant contributions in all vertices
with $n\geq 4$. Note that these contributions scale with time as ${\rm
e}^{-3\e}$. Therefore, they decay compared to the higher-point
vertices
induced by non-linear evolution which scale as ${\rm e}^{-2\e}$. For
the sake of the presentation, we will focus on the Gaussian case in the rest of the paper. 

We now turn to the counterterms 
$C_n$. From eq.~(\ref{eq:Wdot}) and in the limit $\eta_0\to -\infty$ they are
found to be time-independent and determined by the
recursion relations,
\be
\label{counter}
\begin{split}
&C_n(\k_1,\ldots,\k_n)=\frac{1}{n}\bigg[\delta^{(3)}\bigg(\sum_{i=1}^n\k_i\bigg)
\int [dq]\,I_{n+1}(\q,\k_1,\ldots,\k_n)\\
&~~~~~~-\sum_{m=2}^n\frac{1}{m!(n-m)!}\sum_\s
I_m\big(\k_{\s(1)},\ldots,\k_{\s(m)}\big)\,
C_{n-m+1}\Big(\sum_{l=1}^m\k_{\s(l)},\k_{\s(m+1)},\ldots,\k_{\s(n)}\Big)\bigg]\,,
\end{split}
\ee
where\footnote{The counterterm $C_0$ which is also
  formally generated according to (\ref{eq:Wdot}) gets absorbed into
  the normalization of the distribution function.} 
$n\geq 1$ and for $n=1$ the sum should be omitted. Comparing this
with the ``tree-level'' weight (\ref{treeact}) we see that the
counterterms are suppressed by the second power of the coupling
constant $g(\e)$. This is precisely the suppression expected for
1-loop contributions. Indeed, we will see that $C_n$ cancel certain UV divergences
of the 1-loop expressions. 

Finally, statistical homogeneity implies that all the
vertices and counterterms are proportional to a $\delta$-function of
the sum of the momenta entering them. We will use prime to denote the
quantities stripped of this $\delta$-function, as has become customary in the
studies of LSS,
\be
\label{eq:tildeGC}
\bar\G_n={\bar\G}'_n(\k_1,\ldots,\k_n)\,\delta^{(3)}\Big(\sum^n_{i=1}
\k_i\Big)\,, \qquad 
C_n= \Ct_n(\k_1,\ldots,\k_n)\, \delta^{(3)}\Big(\sum^n_{i=1} \k_i\Big).
\ee

\subsection{Perturbative expansion}\label{sec:pert}

We want to compute the correlation functions by expanding the
generating functional (\ref{eq:ztfp}) perturbatively in the coupling
constant $g(\e)$. As $g^2$ appears in all expressions multiplied by
the linear power spectrum $\bar P(|\k|)$, the expansion in $g^2$ is
equivalent to the expansion in powers of the initial spectrum
used in SPT. Thus, the two approaches should agree
when comparing the expressions for the correlators at the same fixed order.

The computation is organized by expanding around the Gaussian part of
$\bar\G[\T]$. We write,
\be
\label{eq:ztsptfun}
\begin{split}
Z[J;\e]=\mathcal{N}^{-1}\int [\mathcal{D}\T]\exp 
\bigg\{&-\frac{1}{g^2(\e)}
\sum_{n=2}^{\infty}\int\frac{[dk]^n}{n!}\,\bar\G_n\,
\prod_{j=1}^n\T(\k_j)\\
&-\sum_{n=1}^{\infty}\int\frac{[dk]^n}{n!}\,C_n\,
\prod_{j=1}^n\T(\k_j)
+\int[dk] \T(\k)J(-\k)\bigg\}\\
=\exp \bigg\{\!\!
-\frac{1}{g^2(\e)} \sum_{n=3}^{\infty}\int \frac{[dk]^n}{n!}
&\bar\G_n\prod_{j=1}^n\frac{\delta}{\delta J(-\textbf{k}_j)}
-\sum_{n=1}^{\infty}\int \frac{[dk]^n}{n!}C_n\prod_{j=1}^n
\frac{\delta}{\delta J(-\textbf{k}_j)}
\bigg\}Z^{(2)}[J;\e]\,,
\end{split}
\ee
where 
\be
Z^{(2)}[J;\e]= {\cal N}^{-1} \exp\bigg\{\frac{g^2(\e)}{2}
\int [dk] \,
\bar P(|\k|)\,J(\k)J(-\k)\bigg\}
\ee
is the Gaussian part. Taylor expansion of the exponential in
\eqref{eq:ztsptfun} and use of Wick's theorem generates Feynman diagrams with the propagator
$g^2\bar P(|\k|)$ and vertices $\bar\G_n/g^2$, $n\geq 3$ and $C_n$,
$n\geq 1$. The first building blocks for these diagrams are shown in
Fig.~\ref{fig:feynmanrules}. The Feynman rules are similar to those for a
scalar QFT in a 3-dimensional space with Euclidean signature. Unlike an ordinary QFT,
 the expansion in TSPT contains vertices
with an arbitrary number of legs, and all vertices have non-trivial
momentum dependence. The latter property implies that in position space
the theory is non-local. This non-locality does not lead to any problems in the
perturbative expansion and is present also in SPT. 
In contrast to SPT, time does not flow
along 
the diagrammatic elements, but is taken care of by the time dependence
of the coupling constant.

\begin{figure}
\begin{align*}
\begin{fmffile}{example-ps}
\parbox{90pt}{
\begin{fmfgraph*}(80,80)
\fmfpen{thick}
\fmfleft{l1}
\fmfright{r1}
\fmf{plain,label=${\bf k}$}{l1,r1}
\end{fmfgraph*}}
\end{fmffile}&=~g^2(\e)\bar P(|\k|),\hspace{-2.3cm}&\\
\begin{fmffile}{example-bisp_f}
\parbox{100pt}{
\begin{fmfgraph*}(90,70)
\fmfpen{thick}
\fmfleft{l1,l2}
\fmfright{r1}
\fmf{plain,label=${\bf k}_1$,label.side=right}{l1,b1}
\fmf{plain,label=${\bf k}_2$,label.side=left}{l2,b1}
\fmf{plain,label=${\bf k}_3$}{b1,r1}     
\end{fmfgraph*}}
\end{fmffile}&=-g^{-2}(\e)\frac{\bar\G_3(\k_1,\k_2,\k_3)}{3!},&
\begin{fmffile}{example-C1}
\parbox{80pt}{
\begin{fmfgraph*}(75,75)
\fmfpen{thick}
\fmfleft{l1}
\fmfright{r1}
\fmf{plain,label=${\bf k}$}{l1,r1}
\fmfv{d.sh=cross,d.si=.15w}{l1}
\end{fmfgraph*}}
\end{fmffile}&=-C_1(\k),
\\
\begin{fmffile}{example-trisp_f}
\parbox{100pt}{
\begin{fmfgraph*}(90,70)
\fmfpen{thick}
\fmfleft{l1,l2}
\fmfright{r1,r2}
\fmf{plain,label=${\bf k}_1$}{l1,b1}
\fmf{plain,label=${\bf k}_2$,label.side=right}{l2,b1}
\fmf{plain,label=${\bf k}_4$,l.s=left}{b1,r1}  
\fmf{plain,label=${\bf k}_3$,label.side=left}{b1,r2}  
\end{fmfgraph*}}
\end{fmffile}&=-g^{-2}(\e)\frac{\bar\G_4(\k_1,\k_2,\k_3,\k_4)}{4!},&
~\hspace{0.8cm}
\begin{fmffile}{example-C2}
\parbox{80pt}{
\begin{fmfgraph*}(75,40)
\fmfpen{thick}
\fmfleft{l1}
\fmfright{r1}
\fmf{plain,label=${\bf k}_1$}{l1,v1}
\fmf{plain,label=${\bf k}_2$}{v1,r1}
\fmfv{d.sh=cross,d.si=.15w}{v1}
\end{fmfgraph*}}
\end{fmffile}&=-\frac{C_2(\k_1,\k_2)}{2}
\end{align*}
\caption{Example of TSPT Feynman diagrams.\label{fig:feynmanrules}}
\end{figure}	

It is instructive to consider the tree-level expressions for the 3-
and 4-point correlators. Using the diagrams depicted in
Fig.~\ref{fig:feynmanrules} one obtains,
\begin{align}
\label{eq:amputated}
\langle \T_\e(\k_1)&\T_\e(\k_2)\T_\e(\k_3)\rangle^\mathrm{tree}=
\begin{fmffile}{example-bisp-small}
\parbox{50pt}{
\begin{fmfgraph*}(50,40)
\fmfpen{thick}
\fmfleft{l1,l2}
\fmfright{r1}
\fmf{plain}{l1,b1}
\fmf{plain}{l2,b1}
\fmf{plain}{b1,r1}     
\end{fmfgraph*}}
\end{fmffile}\,=
-g^4(\e)\prod_{i=1}^3 \bar P(|\k_i|) \;\bar\G_3(\k_1,\k_2,\k_3)\,,\\
~\notag
\\
\langle
\T_\e(\k_1)&\T_\e(\k_2)\T_\e(\k_3)\T_\e(\k_4)\rangle^\mathrm{tree}=
\begin{fmffile}{example-trisp-small1}
\parbox{50pt}{
\begin{fmfgraph*}(50,40)
\fmfpen{thick}
\fmfleft{l1,l2}
\fmfright{r1,r2}
\fmf{plain}{l1,b1}
\fmf{plain}{l2,b1}
\fmf{plain}{b1,r1}  
\fmf{plain}{b1,r2}   
\end{fmfgraph*}}
\end{fmffile}
+
\begin{fmffile}{example-trisp-small2}
\parbox{80pt}{
\begin{fmfgraph*}(80,40)
\fmfpen{thick}
\fmfleft{l1,l2}
\fmfright{r1,r2}
\fmf{plain}{l1,b1}
\fmf{plain}{l2,b1}
\fmf{plain}{b1,b2}
\fmf{plain}{b2,r1}  
\fmf{plain}{b2,r2}   
\end{fmfgraph*}}
\end{fmffile}\notag\\
=&g^6(\e)\prod_{i=1}^4\bar P(|\k_i|) 
\;\bigg[-\bar\G_4(\k_1,\k_2,\k_3,\k_4)\notag\\
&+\delta^{(3)}\Big(\sum_{j=1}^4\k_j\Big)
\Big(\bar\G_3'(\k_1,\k_2,-\k_1-\k_2)\,\bar P(|\k_1+\k_2|)\,
\bar\G_3'(\k_1+\k_2,\k_3,\k_4)+\text{perm.}\Big)
\bigg]\,,
\end{align}
where ``perm.'' in the last expression stands for the terms obtained
by the exchange $\k_2\leftrightarrow\k_3$ and $\k_2\leftrightarrow\k_4$.
We observe that $\bar\G_n$ are identified as one-particle-irreducible
(1PI) contributions to the tree-level correlators with amputated
external propagators.  

As already noted above, the counterterms $C_n$ have the same order in
the coupling $g$ as the 1-loop contributions. To understand their
role, consider the 1-loop correction to the average of $\T$,
\be
\label{1ptadple}
\begin{split}
\langle \T_\e(\k)\rangle &= ~
\begin{fmffile}{p1-2}
\parbox{80pt}{
\begin{fmfgraph*}(80,30)
\fmfpen{thick}
\fmfkeep{1loop_2}
\fmfleft{l1}
\fmfright{r1}
\fmfv{d.sh=circle,d.filled=full,d.si=.01w,l.a=135,l.d=.08w}{v1}
\fmfv{d.sh=circle,d.filled=full,d.si=.001w,label=$ $,l.a=45,l.d=.05w}{v2}
\fmf{plain,label=$ $}{l1,v1}
\fmf{phantom,label=$ $}{v2,r1}
\fmf{plain,left=1.0,tension=0.5,label=${\bf q}$}{v1,v2}
\fmf{plain,left=1.0,tension=0.5,label=$ $,l.side=left}{v2,v1}
\fmfposition
\end{fmfgraph*}}
\end{fmffile} 
\!\!\!\!\!\!\!\!\!+~~
\begin{fmffile}{p1-1}
\parbox{50pt}{
\begin{fmfgraph*}(50,1)
\fmfpen{thick}
\fmfleft{l1}
\fmfright{r1}
\fmf{plain,label=$ $}{l1,v1}
\fmf{phantom,label=$ $}{v1,r1}
\fmfv{d.sh=cross,d.si=.25w,l.a=90,l.d=.2w}{v1}
\end{fmfgraph*}}
\end{fmffile}  \\
  &=-g^2(\e)\bar P(|\k|)\delta^{(3)}(\k)\bigg[
\frac{1}{2}\int [dq]\bar\G_3'(\k,\q,-\q)\,\bar P(|\q|)+C_1'(\k)\bigg] \,.
\end{split}
\ee
Using the expressions (\ref{Wvert3}), (\ref{eq:C1}) for $\bar\G_3'$
and $C_1'$ we find that the combination in the square bracket
vanishes, provided $I_2(\q,-\q)=0$. The latter condition is satisfied
in the case of fluid dynamics and ZA (cf. eqs.~(\ref{eq:IZA2}),
(\ref{identificED2}), (\ref{eq:K2ED})). We conclude that $C_1$ cancels the unphysical
{\it tadpole} graph. More generally, by inspection of other 1-loop diagrams
one finds that $C_n$ cancel certain UV divergent contributions. A detailed study of this cancellation
for the  $\T$-power spectrum at 1-loop can be found in
appendix~\ref{app:1loop}. Diagrammatically, this can be expressed as
\be
\begin{split}
\label{diagr1loop0}
&P^\mathrm{1-loop}_{\T\T}(\eta;|\k|)= 
~
\begin{fmffile}{4-vertex-ex1}
\parbox{80pt}{
\begin{fmfgraph*}(80,35)
\fmfpen{thick}
\fmfkeep{1loop}
\fmfleft{l1,l2}
\fmfright{r1,r2}
\fmflabel{$\bar\G_4$}{b1}
\fmf{plain,label=${\bf k}$,l.s=right}{l1,b1}
\fmf{plain,label=${\bf k}$,l.s=right}{b1,r1}
\fmf{phantom}{l2,u1}
\fmf{phantom}{u2,r2}
\fmf{plain,left=0.5,tension=0.01,l.side=left}{b1,u1}
\fmf{plain,left=0.5,tension=2,label=$\bf q$}{u1,u2}
\fmf{plain,right=0.5,tension=0.01}{b1,u2}
\end{fmfgraph*}}
\end{fmffile}
~
+
\quad
\begin{fmffile}{3-vertex-ex1}
\begin{fmfgraph*}(80,1)
\fmfpen{thick}
\fmfkeep{1loop_2}
\fmfleft{l1}
\fmfright{r1}
\fmfv{d.sh=circle,d.filled=full,d.si=.01w,label=$\bar\G_3$,l.a=135,l.d=.05w}{v1}
\fmfv{d.sh=circle,d.filled=full,d.si=.01w,label=$\bar\G_3$,l.a=45,l.d=.05w}{v2}
\fmf{plain,label=${\bf k}$}{l1,v1}
\fmf{plain,label=${\bf k}$}{v2,r1}
\fmf{plain,left=1.0,tension=0.5,label=${\bf q}$}{v1,v2}
\fmf{plain,left=1.0,tension=0.5,label=${\bf q}-{\bf k}$,l.side=left}{v2,v1}
\fmfposition
\end{fmfgraph*}
\end{fmffile}
\quad 
+
\quad
\begin{fmffile}{example-ps01}
\parbox{40pt}{
\begin{fmfgraph*}(40,1)
\fmfpen{thick}
\fmfleft{l1}
\fmfright{r1}
\fmf{plain,label=${\bf k}$}{l1,v1}
\fmf{plain,label=${\bf k}$}{v1,r1}
\fmfv{d.sh=cross,d.si=.25w,label=$C_2$,l.a=90,l.d=.2w}{v1}
\end{fmfgraph*}}
\end{fmffile}  
\end{split}
\ee
	    
\vspace{.6cm}\noindent	    
This calculation allows us to verify explicitly that the total result
for $P_{\T\T}^\mathrm{1-loop}$ in TSPT coincides with the standard SPT
expression, though the contributions of the individual diagrams in
(\ref{diagr1loop0}) are found to be markedly different from those in
SPT. In particular, the {\it daisy} (first term) and {\it fish}
(second term) diagrams in
(\ref{diagr1loop0}) contain extra UV divergences, not present in
SPT. Some of them are canceled between the {\it daisy} and the {\it
  fish}, whereas the rest are canceled by the counterterm $C_2$. We do
not consider the appearance of extra UV divergent contributions as a
drawback of our formalism, because the fluid description of LSS anyway
requires a UV renormalization (see the discussion in the Introduction). Concerning the IR, and in contrast to SPT, all loop diagrams in
(\ref{diagr1loop0}) are manifestly IR-finite. In sec.~\ref{sec:IRsafe} we
will prove that this property holds for all building blocks of the TSPT
expansion.   

To avoid confusion, let us stress that although the $C_n$-vertices act
as counterterms, they differ from generic counterterms of QFT in that
their values cannot be adjusted at will: they 
are fixed unambiguously by the
solution of the Liouville equation. 
They are required to reproduce eventually the SPT result in the
perfect fluid case.  
The UV renormalization of the theory is likely to require additional
counterterms to capture the genuine physical effects of the short
modes.

\subsection{More than one field}\label{sec:multifield}

So far we have described the TSPT framework for a single field
with random initial conditions. In the case of LSS we
usually work with two fields --- the velocity divergence $\T_\e$ and
the density contrast $\delta_\e$. The initial conditions for these
variables are related: they correspond to the adiabatic linear growing
mode. This means that only one of the fields is statistically
independent and can act as the argument of the probability
distribution function. We choose it to be the velocity divergence
$\T_\e$. The density contrast must be expressed in terms of $\T_\e$
using the equations of motion.\footnote{In general, for systems with
  several statistically independent degrees of freedom, the
  probability density and all the fields must be expressed as functions
of the complete set of statistically independent variables.}
In the spirit of perturbation theory this relation can be written as
Taylor series,
\be
\label{eq:psi1}
\delta_\e(\k)\equiv\delta[\T_\e;\e,\k]= \sum_{n=1}^\infty\frac{1}{n!}
\int [dq]^n K_n(\e;\q_1,...,\q_n)\,\delta^{(3)}\Big(\k-\sum_{i=1}^n
\q_i\Big)
\prod_{j=1}^n\T_\e(\q_j)  \,.
\ee
In the next section we will show how to determine the kernels $K_n$
from the equations of fluid mechanics.\footnote{It turns out that in the relevant cases $K_n$ are time-independent, see eqs.~(\ref{eq:Kza}), (\ref{eq:Ked}). For the sake of generality, we presently keep the argument $\e$ in their expressions.}

To compute the correlators involving the density contrast one
generalizes the generating functional by inclusion of a source
$J_\delta$ coupled to $\delta_\e$,
\be
\label{eq:ZJJ}
Z[J,J_\delta;\e]=
\mathcal{N}^{-1}\!\!\int [\mathcal{D}\T]
\exp\bigg\{\!\!-\Gamma[\T;\e]+\int[dk]\T(\k)
J(-\k)+\int[dk]\,\delta[\T;\e,\k]\, J_\delta(-\k)
\bigg\}\,.
\ee
Notice that the sources $J$ and $J_\delta$ enter on different footing. While $J$ couples directly to the `elementary' field $\T$, $J_\delta$ multiplies a series in $\T^n$. This means that in the QFT language $\delta$ should be interpreted as a {\em composite} operator. Variation of (\ref{eq:ZJJ}) with respect to $J_\delta$ produces a set of vertices with multiple legs, which we will denote by a thick dot with an arrow indicating the flow of momentum into or out of the vertex, see Fig.~\ref{fig:composite} .

\begin{figure}
\[
\begin{fmffile}{composite}
\parbox{120pt}{
\begin{fmfgraph*}(110,100)
\fmfpen{thick}
\fmfleft{l1}
\fmfright{r1,r2,r3,r4,r5,r6}
\fmfv{d.sh=circle,d.filled=full,d.si=3thick}{b1}
\fmf{phantom_arrow,tension=10,label=$\k$,l.s=left}{l1,b1}
\fmf{plain}{b1,v6}
\fmf{plain,label=$\q_1$,l.s=left}{v6,r6}
\fmf{plain}{b1,v5}
\fmf{plain,label=$\q_2$,l.s=left,l.d=4}{v5,r5}
\fmf{phantom}{b1,v4}
\fmf{phantom,label=$\bullet$,l.s=right,l.d=0}{v4,r4}
\fmf{phantom}{b1,v3}
\fmf{phantom,label=$\bullet$,l.s=right,l.d=0}{v3,r3}
\fmf{phantom}{b1,v2}
\fmf{phantom,label=$\bullet$,l.s=right,l.d=0}{v2,r2}
\fmf{plain}{b1,v1}
\fmf{plain,label=$\q_n$,l.s=right}{v1,r1}
\end{fmfgraph*}}
\end{fmffile}=\frac{K_n(\e;\q_1,\ldots,\q_n)}{n!}\,\delta^{(3)}\Big(\k-\sum_{i=1}^n\q_i\Big)
\]
\caption{Vertices appearing in the expansion of $\delta$ treated as a composite operator.\label{fig:composite}}
\end{figure}

Let us illustrate this point with the power spectrum for the density field,
 \be
 \langle \delta_\e(\k) \delta_\e(\k')\rangle=
 \left.\frac{\delta^2Z[J,J_\delta;\e]}{\delta J_\delta(-\k)\delta J_\delta(-\k')}\right|_{J,J_\delta=0}.
\ee 
We set $K_1=1$, which corresponds to the equality between $\delta_\e$ and $ \T_\e$ at the linear level; this is consistent with the adiabatic initial conditions for LSS. Then, the power spectrum of $\delta$ at a fixed loop order is given by the same diagrams as the power spectrum of $\T$ plus a number of extra diagrams containing one or two of the vertices from Fig.~\ref{fig:composite}  with $n\geq 2$. Namely, for one-loop corrections we have,
\be
\label{eq:dPSfromth}
P_{\delta\delta}^\mathrm{1-loop}(\eta;|\k|)=P_{\T\T}^\mathrm{1-loop}(\eta;|\k|)+\hat P^\mathrm{1-loop}_{\delta\delta}(\eta ;|\k|),
\ee
where $P_{\T\T}^\mathrm{1-loop}$ is given by eq.~(\ref{diagr1loop0}), whereas the extra contribution has three diagrams 
\be
\label{eq:Ddelta}
\hat P^\mathrm{1-loop}_{\delta\delta}(\eta;|\k|)=  
\quad
\begin{fmffile}{cs1toy}
\parbox{90pt}{
\begin{fmfgraph*}(80,60)
\fmfpen{thick}
\fmfleft{l1}
\fmfright{r1}
\fmfv{d.sh=circle,d.filled=full,d.si=3thick,label=$K_2$,l.a=120,l.d=.05w}{v1}
\fmfv{d.sh=circle,d.filled=full,d.si=.01w,label=$\bar\G_3$,l.a=60,l.d=.05w}{v2}
\fmf{phantom_arrow,tension=3,label=$\k$,l.s=right}{l1,v1}
\fmf{plain,tension=2,label=$\bf k$,l.s=right}{v2,r1}
\fmf{plain,left=0.85,tension=0.5,label=$\q$}{v1,v2}
\fmf{plain,left=0.85,tension=0.5,label=${\bf q}-{\bf k}$,l.side=left}{v2,v1}
\end{fmfgraph*}}
\end{fmffile}
+~~	  
\begin{fmffile}{cs2toy}
\parbox{90pt}{
\begin{fmfgraph*}(80,60)\vspace{1cm}
\fmfpen{thick}
\fmfleft{l1}
\fmfright{r1}
\fmfv{decor.shape=circle,d.filled=full,d.si=3thick,label=$K_2$,l.a=120,l.d=.05w}{v1}
\fmfv{decor.shape=circle,d.filled=full,d.si=3thick,label=$K_2$,l.a=50,l.d=.05w}{v2}
\fmf{phantom_arrow,tension=2.3,label=$\k$,l.s=right}{l1,v1}
\fmf{phantom_arrow,tension=2.3,label=$\k$,l.s=right}{v2,r1}
\fmf{plain,left=0.85,tension=0.5,label=${\bf q}$}{v1,v2}
\fmf{plain,left=0.85,tension=0.5,label=${\bf q}-{\bf k}$,l.side=left}{v2,v1}
\end{fmfgraph*}}
\end{fmffile}
+	
~~	   	    
\begin{fmffile}{cs3toy}
\parbox{80pt}{
\begin{fmfgraph*}(70,60)
\fmfpen{thick}
\fmfleft{l1,l2,l3}
\fmfright{r1,r2,r3}
\fmf{phantom_arrow,tension=2,label=$\k$,l.s=right}{l2,b1}
\fmf{plain,label=$\bf k$,l.s=right}{b1,r2}
\fmf{plain,left=0.5,tension=0.01,l.side=left}{b1,u1}
\fmf{plain,left=0.1,tension=3,label=$\bf q$}{u1,u2}
\fmf{plain,right=0.5,tension=0.01}{b1,u2}
\fmf{phantom,tension=1}{u1,r3}
\fmf{phantom,tension=0.3}{u1,l3}
\fmfv{decor.shape=circle,decor.filled=full,decor.size=3thick,label=$K_3$,l.a=120,l.d=.07w}{b1}
\end{fmfgraph*}}
\end{fmffile} 
\ee

\noindent 
These diagrams are evaluated in appendix~\ref{app:1loop}. As for the $\T$ case,  we show that 
$P_{\delta\delta}^\mathrm{1-loop}$ coincides with the SPT one-loop result.

\section{Application to Zel'dovich approximation and exact
  dynamics}\label{Sec:applic} 

In this section we apply the general formalism developed above to the
dark matter fluid. We consider the Zel'dovich approximation and exact
dynamics, which are defined by their corresponding 
evolution kernels $I_n$ and $K_n$.

\subsection{Zel'dovich approximation}\label{sec:zel}

In the ZA, the equation \eqref{psidotED2} is modified by promoting the
linear relation between the fields $\delta$ and $\T$ to the non-linear
level. Namely, one substitutes $\delta=\T$ in the l.h.s. of
\eqref{psidotED2}, which allows to decouple the equation for $\T$
completely from the density field, 
\be
\label{psidot}
\dot \T_\e(\k)- \T_\e(\k)=
\int[dq]^2\,\delta^{(3)}(\k-\q_1-\q_2)\,\b(\q_1,\q_2)\T_\e(\q_1)\T_\e(\q_2)\,. 
\ee
Comparing this equation to \eqref{eq:eomp}, \eqref{eq:ik} one identifies, 
\bseq
\label{eq:IZA}
\begin{align}
&I^{ZA}_1=1,\\
&I^{ZA}_2(\k_1,\k_2)=2\beta(\k_1,\k_2), \label{eq:IZA2}
\\
\label{eq:IZA3}
&I^{ZA}_n=0~,~~~~n\geq 3\;.
\end{align}
\eseq
In consequence, the recursion relation \eqref{solGaus2} for the
vertices 
takes a simple form,
\be
\label{recurG}
\bar\G^{ZA}_{n}(\k_1,...,\k_{n})=- \frac{2}{n-2}
\sum_{1\leq i<j\leq n}\b(\k_{i},\k_{j})\,
\bar\G^{ZA}_{n-1}(\k_{i}+\k_j,\k_1,\ldots,\check \k_i,\ldots,\check \k_j,\ldots,\k_n)\,,
\ee
where the notations $\check \k_i$, $\check \k_j$ mean that the kernel
$\bar\G^{ZA}_{n-1}$ does {\em not} have these momenta among its
arguments. The seed member of the recursion $\bar\G_2$ is set
by eq.~(\ref{initGaus1}).

We now turn to the counterterms $C_n$. Substituting the expressions
(\ref{eq:IZA}) into 
eqs.~(\ref{eq:C12}) from appendix~\ref{app:1loop} we obtain,
\begin{subequations}
\begin{align}
&C^{ZA}_1(\k)=2\,\delta^{(3)}(\k)\int [dq]\;\b(\q,\k)\,, \\
\label{CnZA}
&C^{ZA}_2(\k_1,\k_2)= -2\,\delta^{(3)}(\k_1+\k_2)\,\b(\k_1,\k_2)\int
[dq]\,\b(\q,\k_1+\k_2)\,.
\end{align}
\end{subequations}
Note that the asymptotic behavior $\b(\k,\q)=O(q/k)$ at $q\to\infty$
implies that the integrals on the r.h.s. are UV divergent. In
principle, they should be regularized by introduction of a UV
cutoff. We won't need the details of this regularization: in the
actual computations the divergences cancel when summed with the loop
contributions, see appendix~\ref{app:1loop}. 
In ZA, the situation is actually even simpler. Using the property
\be
\label{betazero}
\lim_{p \to 0}\b(\q,{\bf p})\b(\k,-\k+{\bf p})=
\lim_{p\to 0}\frac{(\q\cdot{\bf p})}{p^2}\frac{p^2}{k^2} =0 \,,
\ee
we infer that $C_2^{ZA}$ vanishes. Then, the recursion
relation (\ref{counter}) together with 
eq.~(\ref{eq:IZA3}) imply that all $C_n$ with $n>2$ vanish as
well. We conclude that in ZA only the first counterterm $C_1$ is non-zero. As
discussed in sec.~\ref{sec:pert}, its role is precisely to cancel the
{\it tadpole} contributions. Thus, in ZA one
can  forget both about the counterterms and tadpoles in the
diagrammatic expansion.

\subsubsection{The density field}\label{sec:densityZA}

The vertices $\bar\G_n$ are
sufficient to compute the correlators of the velocity dispersion $\T$
in ZA. For the correlators of the density field $\delta$ we need to know also
the kernels $K_n$. Substituting the representation (\ref{eq:psi1})
into the continuity equation (\ref{psidotED1}) we find that these
satisfy a system of linear differential equations,
\begin{subequations}
\label{Kn}
\begin{align}
&\dot K^{ZA}_1(\e;\k)+K^{ZA}(\e;\k)=1\,, \\
\label{KZA2}
&\dot{K}^{ZA}_2(\e;\k_1,\k_2)+2K^{ZA}_2(\e;\k_1,\k_2)=
\a(\k_1,\k_2)K_1^{ZA}(\e;\k_2)+\a(\k_2,\k_1)K_1^{ZA}(\e;\k_1)\notag\\
&~~~~~~~~~~~~~~~~~~~~~~~~~~~~~~~~~~~~~~~~~~~~~~~
-2\b(\k_1,\k_2)
K_1^{ZA}(\e;\k_1+\k_2)\,,\\
&\dot{K}^{ZA}_n(\e;\k_1,\ldots,\k_n)+nK^{ZA}_n(\e;\k_1,\ldots,\k_n)\notag\\
&~~~~~~~~~~~~~~~~~~~~~~~~~
+2\!\!\sum_{1\leq i<j\leq n}\!\! \b(\k_{i},\k_{j})\,
K^{ZA}_{n-1}(\e;\k_i+\k_j,\k_{1},\ldots,\check\k_i,\ldots,\check\k_j,\ldots,\k_n)
\notag\\
&~~~~~~~~~~~~~~~~~~~~~~~~~
=\sum_{i=1}^n\a\Big(\k_i,\!\sum_{1\leq j\leq n,\,j\neq i}\!\!
\k_j\Big)\,
K^{ZA}_{n-1}(\e;\k_1,\ldots,\check{\k}_i,\ldots,\k_n)~,~~~~~~~~n\geq 3\,.
\end{align}
\end{subequations}
For the adiabatic mode $\delta_\e=\T_\e$
at $\e\to-\infty$. This corresponds to the initial conditions on the kernels,
\be
\label{eq:Kzainit}
\lim_{\eta\to-\infty} K_1(\e)=1~,~~~~ 
\lim_{\eta\to-\infty} K_n(\eta)\,{\rm e}^{(n-1)\e}=0~,~~~n\geq 2\,.
\ee
The solution of \eqref{Kn} that satisfy these initial conditions is
time-independent and is given by the recursions formulas, 
\bseq 
\label{eq:Kza}
\begin{align}
\label{eq:Kza1}
&K_1^{ZA}=1\;,\\
\label{eq:Kza2}
&K^{ZA}_2(\k_1,\k_2)= 1-\frac{(\k_1\cdot \k_2)^2}{k_1^2k_2^2}
\equiv \sin^2(\k_1,\k_2)\,,\\
&K^{ZA}_n(\k_1,\ldots,\k_n)=\frac{1}{n}\bigg[
\sum_{i=1}^n\a\Big(\k_i,\!\sum_{1\leq j\leq n,\,j\neq i}\!\!
\k_j\Big)\,
K^{ZA}_{n-1}(\k_1,\ldots,\check{\k}_i,\ldots,\k_n)\notag
\\
&~~~~~~~~~~~~~~~~~~~~~~~
-2\!\!\sum_{1\leq i<j\leq n}\!\! \b(\k_{i},\k_{j})\,
K^{ZA}_{n-1}(\k_i+\k_j,\k_{1},\ldots,\check\k_i,\ldots,\check\k_j,\ldots,\k_n)\bigg]
\;,~~~~~n\geq 3\,.
\label{KnZA}
\end{align}
\eseq
To sum up, eqs.~(\ref{recurG}), (\ref{eq:Kza}) determine all ingredients
of the TSPT diagrammatic expansion within Zel'dovich approximation.

\subsection{Exact dynamics}\label{sec:ED}

We now repeat the above derivation for exact dynamics. In this case the equations for the kernels $I_n$ cannot be decoupled and must be solved together with the equations for $K_n$. We proceed as follows. First, we use eq.~(\ref{psidotED2}) to express $I_n$ in terms of $K_n$. Substituting the expansion (\ref{eq:psi1}) we obtain,\footnote{Recall that we work in the approximation $\Omega_m/f^2=1$.}
\bseq
\label{identificED}
\begin{align}
&I_1(\e;\k)=-\frac{1}{2}+\frac{3}{2}K_1(\e;\k)\;,\\
\label{identificED2}
&I_2(\e;\k_1,\k_2)=2\b(\k_1,\k_2)+\frac{3}{2}K_2(\e;\k_1,\k_2)\;,\\
&I_n(\e;\k_1,\ldots,\k_n)=\frac{3}{2}K_n(\e;\k_1,\ldots,\k_n)~,~~~~n\geq 3\;.
\end{align}
\eseq
Next, we insert the expansions (\ref{eq:ik}), (\ref{eq:psi1}) into (\ref{psidotED1}). After a straightforward calculation one arrives to a chain of differential equations,
\bseq
\label{eq:KdotED}
\begin{align}
\dot K_1(\e;\k)+I_1(\e;\k&)K_1(\e;\k)=1\;,\\
\dot{K}_n(\e;\k_1,...,\k_n)+&\sum_{m=1}^n\frac{1}{m!(n-m)!}
\sum_\s I_m(\e;\k_{\s(1)},\ldots,\k_{\s(m)})\notag\\
&\times
K_{n-m+1}\Big(\sum_{l=1}^m \k_{\s(l)},\k_{\s(m+1)},\ldots,\k_{\s(n)}\Big)\notag\\
&=\sum_{i=1}^n \a\Big(\k_i,\!\sum_{1\leq j\leq n,j\neq i}\!\! \k_j\Big)\, K_{n-1}(\k_1,\ldots,\check{\k}_i,\ldots,\k_n)\;,~~~~n\geq 2\;.
\end{align}
\eseq
Using eqs.~(\ref{identificED}) we obtain a closed system for the kernels $K_n$.
Note that the equation for the $n-$th kernel depends only on $K_m$ with $m\leq n$, so the system can be solved exactly. The adiabatic initial conditions (\ref{eq:Kzainit}) uniquely fix the solution, which as in ZA case, is found to be time-independent. We obtain the recursion relations,
\begin{subequations}
\label{eq:Ked}
\begin{align}
\label{Ked1}
&K_1=1\;,\\
\label{eq:K2ED}
&K_2(\k_1,\k_2)=\frac{4}{7}\sin^2(\k_1,\k_2)\,,\\
&K_n(\k_1,\ldots,\k_n)=\frac{2}{2n+3}\bigg[
\sum_{i=1}^{n}\a\Big(\k_i,\!\sum_{1\leq j\leq n,j\neq i} \!\!\k_j\Big)\,
K_{n-1}(\k_1,\ldots,\check{\k}_i,\ldots,\k_{n})\nonumber\\
&-\sum_{1\leq i<j \leq n}\bigg(2\b(\k_i,\k_j)+\frac{3}{2}K_2(\k_i,\k_j)\bigg)\,
K_{n-1}(\k_i+\k_j,\k_1,\ldots,\check\k_i,\ldots,\check\k_j,\ldots,\k_n)\nonumber\\
&-\frac{3}{2} \sum_{m=3}^{n-1}\frac{1}{m!(n-m)!}
\sum_{\s}
K_{m}\big(\k_{\s(1)},\ldots,\k_{\s(m)}\big)\,K_{n-m+1}\Big(\sum_{l=1}^m\k_{\s(l)},\k_{\s(m+1)},\ldots,\k_{\s(n)}\Big)
\Bigg] \,,\notag\\
&\qquad\qquad\qquad\qquad\qquad\qquad\qquad\qquad
\qquad\qquad\qquad\qquad\qquad\qquad\qquad\qquad
n\geq 3\;.
\label{Knsol}
\end{align}
\end{subequations}
The kernels $I_n$ are then retrieved from (\ref{identificED}). Note that (\ref{Ked1}) implies $I_1=1$, justifying the assumption  in eq.~\eqref{eq:I1}. 
Although the recursion relation (\ref{Knsol}) becomes complicated at higher $n$, this does not pose an obstruction for practical computations, which involve only a few
 kernels. For example, to evaluate 1-loop (2-loop) corrections to the power-spectrum one needs the kernels up to $K_3$ ($K_5$).
It is worth mentioning also that if one knows already the SPT kernels, the $K_n$ kernels can be directly found from them, 
see appendix~\ref{app:spttotspt}. 

Once the kernels $K_n$, $I_n$ are known, it is straightforward to construct the vertices $\bar\G_n$ and the counterterms $C_n$ using the general expressions (\ref{solGaus2}), (\ref{counter}). For future reference, let us single out the term containing $I_2$ on the r.h.s. of eq.~(\ref{solGaus2}),
\begin{align}
\nonumber
&\bar\G_n(\k_1,...,\k_n)\!=\!\frac{-1}{n-2} 
\!\sum_{1\leq i<j\leq n}\!\!
\bigg(\!2\beta(\k_i,\k_j)\!+\!\frac{3}{2}K_2(\k_i,\k_j)\!\bigg)
\bar\G_{n-1}(\k_i\!+\!\k_j,\k_1,...,\check\k_i,...,\check\k_j,...,\k_n)
\\
&-\frac{3}{2(n-2)} \sum_{m=3}^{n-1}
\frac{1}{m!(n-m)!}
\sum_{\s}
K_{m}\big(\k_{\s(1)},...,\k_{\s(m)}\big)
\bar\G_{n-m+1}\Big(\sum_{l=1}^m \k_{\s(l)},\k_{\s(m+1)},...,\k_{\s(n)}\Big).
\label{Gned}
\end{align}
Finally, for the first two counterterms we have, 
\begin{subequations}
\begin{align}
&C_1(\k)= \delta^{(3)}(\k)\int [dq]\left(2\b(\q,\k)+\frac{3}{2}K_2(\q,\k)\right)\,,\\
&C_2(\k_1,\k_2)=\frac{3}{4}\delta^{(3)}(\k_1+\k_2)\int [dq]\,K_3(\q,\k_1,\k_2)\,,
\end{align}
\end{subequations}
where we used the general formulas (\ref{eq:C12}) and the property (\ref{eq:cond_1}) that can be easily verified using (\ref{identificED2}), (\ref{eq:K2ED}). We observe that, unlike ZA, the counterterm $C_2$ does not vanish. The same is also true for higher counterterms. They must be properly taken into account in the loop computations.

\section{Soft limits and infrared safety}\label{sec:IRsafe}

Even if the equal time correlators are the same in TSPT and SPT, the intermediate quantities required to compute them are very different. 
As pointed out in the Introduction, in SPT individual diagrams
contain unphysical singularities at low-momenta that cancel only when 
all the diagrams are added up. This complicates the calculations at high loop and hampers the development of diagrammatic resummation techniques which would be desirable to correctly capture the physical effects of IR modes. We show in this section that this problem is absent in TSPT where the individual elements are already IR-safe. 

\subsection{Vertices with soft momenta}\label{sec:5.1}

We want to show that the vertices $\bar\G_n$, $C_n$ and $K_n$
appearing in the TSPT diagrammatic expansion are bounded at finite
values of their arguments\footnote{\label{foot:pq} More precisely, we
  will show that these functions do not have any poles associated to
  the dynamical coupling between the hard and soft modes. The vertices
$\bar\G_n$ (and only them) being inversely proportional to the linear power
spectrum, may, in principle, diverge if the latter vanishes at some
values of the momentum. In the real world this happens at low momenta
where $\bar P(|\q|)\propto q$. However, these divergences cancel in the
{\em individual} TSPT diagrams where the vertices are multiplied by positive
powers of the propagator proportional to $\bar P(|\q|)$. Furthermore, we show
in Appendix~\ref{app:initialPS} that $\bar\G_n$ are finite if at least
two of their arguments are hard and the power spectrum behaves at
small $q$ as $P(|\q|)\propto q^\nu$ with $\nu\leq 2$.}. 
Then, the loop integrals will be free from any IR divergences 
as long as the power spectrum, 
which plays the role of the propagator, 
behaves as\footnote{Recall that
individual loop integrals in SPT do not have IR divergences if $\nu > -1$ \cite{Jain:1995kx}.}  
$\bar P(|\q|) \propto q^{\nu}$ with $\nu >-3$  at $\q\to 0$. 
We  consider ZA and ED in parallel. Recall that the {\em leading} IR behavior in these two cases is identical \cite{Scoccimarro:1995if,Tassev:2011ac}.

Let us first discuss the $K_n$ vertices. One observes that $K_2$ given by (\ref{eq:Kza2}) or (\ref{eq:K2ED}) is manifestly bounded. The proof proceeds by induction. Assume that all $K_m$ with $m<n$ are bounded. From the recursion relations (\ref{KnZA}), (\ref{Knsol}) it is clear that the only potential sources of singularities in $K_n$ are the poles of the kernels $\a$ and $\b$ occurring when either the first argument of $\a$ or one of the two arguments of $\b$ goes to zero, see eqs.~(\ref{alphabeta}). Thus, $K_n$ could potentially have a singularity only if some of the momenta among its arguments vanish. To analyze this limit, let us split all arguments of $K_n$ into  `hard' momenta $\k_1,\ldots,\k_l$ that we keep fixed, and `soft' $\q_1,\ldots,\q_{n-l}$ which are uniformly sent to zero,
\be
\label{softlimit}
\q_s=\epsilon\, \q_s'~,~~~~\epsilon\to 0~,~\q_s'-\text{fixed}\;.
\ee
Focusing on the dangerous terms we obtain,\footnote{Note that we do not write the terms where both arguments of $\a$ or $\b$ are soft, as they are bounded in the limit $\epsilon\to 0$, see eqs.~(\ref{alphabeta}).}
\be
\label{eq:KIRs}
\begin{split}
&K_n(\k_1,..,\k_l,\q_1,...\q_{n-l})
=A_n\bigg[\sum_{s=1}^{n-l}\a\Big(\q_s,\sum_{i=1}^{l}\k_i\Big)\,
K_{n-1}(\k_1,..,\k_l,\q_1,...,\check{\q}_s,...,\q_{n-l})\\
&\qquad\qquad~~~~~
-2\sum_{s=1}^{n-l}\sum_{i=1}^{l}\b(\q_s,\k_i)
K_{n-1}\big(\k_1,..,\k_i+\q_s,...,\k_l,\q_1,...,\check{\q}_s,...,\q_{n-l}\big)\bigg]+{O}(\epsilon^0)\\
&\qquad=A_n\sum_{s=1}^{n-l}\bigg[\a\Big(\q_s,\sum_{i=1}^{l}\k_i\Big)
-2\sum_{i=1}^{l}\b(\q_s,\k_i)\bigg]K_{n-1}(\k_1,..,\k_l,\q_1,...,\check{\q}_s,...,\q_{n-l})+
{O}(\epsilon^0)\,,
\end{split}
\ee
where 
\[
A_n=\begin{cases}
\frac{1}{n}& \text{for ZA},\\
\frac{2}{2n+3}& \text{for ED},
\end{cases}
\]
and $O(\epsilon^0)$ stands for terms that are finite in the limit $\epsilon\to 0$.
The key observation is that the poles of $\a$ and $\b$ at $\q_s\to 0$ cancel in the combination inside  the brackets in the last line of (\ref{eq:KIRs}). Thus, this combination is $O(\epsilon^0)$ and the vertex $K_n$ is IR-finite.

We turn to $\bar\G_n$. The proof again goes by induction. $\bar\G_2$
is given by eq.~(\ref{initGaus1}) and is bounded if the power spectrum
does not vanish. In the realistic cosmology this condition is formally
violated at $q\to 0$ where $\bar P(|\q|)$ behaves linearly. However, as
explained in the footnote~\ref{foot:pq}, this does not pose a threat
to the IR-safety. For the sake of the argument, we are going to assume
that the power spectrum is bounded from below, being concerned only
with those divergences that might arise from the dynamical coupling
between the hard and soft modes. The generalaization of the proof to
the case when the power spectrum behaves as $\bar P(|\q|)\propto q^\nu$
with $\nu\leq 2$ at $q\to 0$ is given in Appendix~\ref{app:initialPS}.

Assume that all $\bar\G_m$ with $m<n$ have been
already shown to be finite. Then, the only contributions in the
recursion relations (\ref{recurG}), (\ref{Gned}) that could induce
singularities of $\bar\G_n$ are the terms containing the kernels
$\b$. Note that they are identical in ZA and ED. Splitting again the
arguments of $\bar\G_n$ into hard and soft, with soft momenta $\q_s$
going uniformly to zero as in (\ref{softlimit}), we isolate the
dangerous part,
\be
\label{GIRsafe}
\bar\G_n(\k_1,...,\k_l,\q_1,...,\q_{n-l})
=\!\frac{-2}{n-2}\sum_{s=1}^{n-l}\!\bigg[\!\sum_{i=1}^l\b(\q_s,\k_i)\!\bigg]
\,\bar\G_{n-1}(\k_1,...,\k_l,\q_1,...,\check\q_i,...,\q_{n-l})+O(\epsilon^0)
\ee
For the sum in brackets we have,
\[
\sum_{i=1}^l\b(\q_s,\k_i)
=\frac{\big(\q_s\cdot\sum_{i=1}^l\k_i\big)}{2q_s^2}+O(\epsilon^0)\;.
\]
Recall now that, due to the momentum conservation, the sum of all
momenta entering into the vertex $\bar\G_n$ must be zero, see
eqs.~(\ref{eq:tildeGC}). This implies,
\be
\sum_{i=1}^l\k_i=-\sum_{s=1}^{n-l}\q_s=O(\epsilon)
~~~~\Longrightarrow ~~~~
\sum_{i=1}^l\b(\q_s,\k_i)=O(\epsilon^0)\;.
\ee
One concludes that all poles on the r.h.s. of (\ref{GIRsafe}) cancel,
and $\bar\G_n$ is also IR-safe.

Finally, this argument can be repeated essentially without changes to
demonstrate the finiteness of $C_n$ using the recursion relation
(\ref{counter}).

\subsection{Relation to the equivalence principle}

The IR-safety of equal-time correlators is known to be closely related
to the symmetry of LSS dynamics \cite{Scoccimarro:1995if} that can be
traced back to the equivalence principle
\cite{Creminelli:2013mca}. This symmetry has been used to derive the
consistency conditions for the IR structure of correlation functions
that do not rely on any specifics of the fluid approximation  
\cite{Kehagias:2013yd,Peloso:2013zw,Creminelli:2013mca,Horn:2014rta}. Here
we explore the implications of this symmetry for TSPT. This approach is
similar to that of \cite{Goldberger:2013rsa} where it was applied to
derive relations between inflationary correlators. Treating
the logarithm of the probability distribution $\G[\T;\e]$ as an
`effective action' of a 3-dimensional Euclidean QFT, we require
it to be invariant under the symmetry transformations and derive the
corresponding conditions on the vertices $\G^{tot}_n$ (Ward
identities). We do not rely on any particular form of the equations of
motion in this section, so our results will be valid for any system
satisfying the relevant symmetry.

Consider the coordinate transformation, 
\bseq
\label{eq:equiv}
\be
\label{eq:equivp}
 \eta \mapsto \eta\,,\quad~~ {\bf x}\mapsto \tilde{\bf x}= {\bf x}-{\rm e}^\e
 \nabla \Phi_L({\bf x}) \,,
\ee
where the function $ \Phi_L({\bf x})$ describes a long-wavelength perturbation. Accordingly, its Fourier transform $\Phi_L(\k)$ has support only at low momenta 
$k<1/L$. The density contrast transforms as a scalar under (\ref{eq:equivp}), while the velocity divergence acquires an inhomogeneous piece due to the time dependence of the coordinate shift,
\be
\label{eq:traf}
\tilde \delta_\e(\tilde {\bf x})= \delta_\e({\bf x})~,~~~~~
\tilde\T_\e(\tilde {\bf x})= \T_\e({\bf x})+{\rm e}^\e\nabla^2\Phi_L({\bf x})\;.
\ee
\eseq
We will be eventually interested in the limit $L \to \infty$, assuming a constant limiting value for the gradient of $\Phi_L$,
\be
\label{limit}
\nabla\Phi_L({\bf x})\to {\bf a}~~~~
\Longleftrightarrow~~~~ \k\Phi_L(\k)\to-i{\bf a}\,\delta^{(3)}(\k)~,
~~~~\text{at}~L\to\infty\;.
\ee
In this limit one can write the transformations (\ref{eq:traf}) in  Fourier space as,
\bseq
\label{eqFour}
\begin{align}
\label{eqFour1}
&\tilde\delta_\e(\k)=\delta_\e(\k){\rm e}^{i({\bf a\cdot k}){\rm e}^\e}\big(1+O(1/L)\big)\;,\\
\label{eqFour2}
&\tilde\T_\e(\k)=\big(\T_\e(\k)-{\rm e}^\e k^2\Phi_L(\k)\big){\rm e}^{i({\bf a\cdot k}){\rm e}^\e}\big(1+O(1/L)\big).\
\end{align}
\eseq
Assume that the original fields $\delta_\e(\k)$, $\T_\e(\k)$ are solutions of the hydrodynamic equations (\ref{psidotED}). Then it is straightforward to check that 
the transformed fields (\ref{eqFour}) also satisfy these equations 
up to terms that vanish as $O(1/L)$. Thus, in the limit (\ref{limit}) the transformations (\ref{eqFour}) become a symmetry of the equations of motion\footnote{Actually, the symmetry group of eqs.~(\ref{psidotED}) is much broader \cite{Kehagias:2013yd,Horn:2014rta}. They are invariant in the limit (\ref{limit}) under the transformations,
\[
\tilde\delta_\e(\k)=\delta_\e(\k){\rm e}^{i({\bf a\cdot k})\xi(\e)}\;,
~~~~~
\tilde\T_\e(\k)=\big(\T_\e(\k)-\dot\xi(\e) k^2\Phi_L(\k)\big){\rm e}^{i({\bf a\cdot k})\xi(\e)}\;,
\]
with an arbitrary function $\xi(\e)$. However, only the time dependence used in (\ref{eqFour}) can be embedded into the full generally relativistic description \cite{Creminelli:2013mca,Horn:2014rta}.}. 
Importantly, this invariance crucially relies on the presence of the inhomogeneous piece in the $\T$-transformation (\ref{eqFour2}). Although this term vanishes in the limit (\ref{limit}), it gets multiplied in the equations of motion by the kernels $\a$ and $\b$ that have poles at low momenta. These poles cancel one factor of $k$ in  $k^2\Phi_L(\k)$, 
which leads to a finite contributions at $L\to\infty$. It should be also stressed that we have not assumed the gradient of the long mode to be small, so the 
invariance holds to arbitrary order in ${\bf a}$.

As discussed in \cite{Creminelli:2013mca,Horn:2014rta}, the transformation (\ref{eq:equiv}) corresponds to superimposing a long-wavelength adiabatic growing mode on top of the original perturbation. The function $\Phi_L({\bf x})$ is proportional to the initial value of the long-mode gravitational potential. The invariance of the dynamics in the limit (\ref{limit}) then follows from the equivalence principle: the transformation (\ref{eq:equiv}) describes the free fall of the short-scale perturbation in the gravitational field of the long mode, which does not affect the local physics. Therefore, this invariance is valid beyond the fluid approximation and holds for any theory obeying the equivalence principle.

The above interpretation implies that the probability to find the transformed field $\tilde\T_\e(\k)$ in the statistical ensemble is equal to the probability of finding the original field $\T_\e(\k)$ times the probability to find the long mode\footnote{Here we use that the initial distribution is Gaussian, so that the long and short modes are statistically independent.}. This leads to the relation,
\be
\label{eqw1}
\W[\tilde \T_\e;\eta]=\W[\T_\e,\eta]+\W[\T_{\e,L};\e],
\ee
where $\T_{\e,L}(\k)=-{\rm e}^\e k^2\Phi_L(\k){\rm e}^{i({\bf a\cdot k}){\rm e}^\e}$. The long mode is in the linear regime, implying that its statistical weight is essentially Gaussian (see eqs.~(\ref{initGaus1}), (\ref{solGaus1})), 
\[
\G[\T_{\e,L};\e]=\frac{{\rm e}^{-2\e}}{2}\int[dk]
\frac{\T_{\e,L}(\k)\T_{\e,L}(-\k)}{\bar P(|\k|)}=
\frac{1}{2}\int[dk]
\frac{k^4\Phi_L(\k)\Phi_L(-\k)}{\bar P(|\k|)}
\;,
\]
As expected, it does not depend on time. Moreover, unless $\bar P(|\k|)$ vanishes at $k\to 0$ as $k^2$ or faster, we have,
\[
\lim_{L\to\infty}\G[\T_{\e,L};\e]=0\;.
\] 
Comparing with (\ref{eqw1}) we obtain that the weight $\G$ must be invariant in the limit (\ref{limit}),
\be
\label{eqw2}
\lim_{L\to\infty}\G[\tilde\T_{\e};\e]=\G[\T_{\e};\e]\;.
\ee
This condition is a consequence of equivalence principle and initial Gaussian statistics.

Using the power-series representation for $\G$ yields a set of equations in the limit (\ref{limit}),
\be
\lim_{L\to\infty}\int [dq]^{n-l}\,\G_n^{tot}(\e;\k_1,\ldots,\k_l,\q_1,\ldots,\q_{n-l})
\prod_{s=1}^{n-l}\big(q_s^2\Phi_L(\q_s)\big)=0
\ee
for all $1\leq l\leq n-1$. This translates into the following conditions on the vertices,
\be 
\label{eqGfin}
\lim_{\epsilon\to 0} \epsilon^{n-l}\, \Wn_n(\e;\k_1,\ldots,\k_l,\epsilon\q_1,\ldots,\epsilon\q_{n-l})=0~,~~~~1\leq l\leq n-1\;.
\ee 
These conditions imply that the vertices $\G_n^{tot}$ cannot have poles of order $(n-l)$ or higher when $(n-l)$ of their momenta become soft. However, they do not forbid, in principle, poles of lower orders. In this respect, they are weaker than the perturbative result of sec.~\ref{sec:5.1} which states that $\G_n^{tot}$ do not have any singularities whatsoever.  They are, however, more powerful in the sense that they are valid beyond the ideal fluid approximation.

As the last remark, we note that the equivalence principle also constrains the IR properties of the vertices $K_n$. Substituting the transformation (\ref{eqFour}) into the relation (\ref{eq:psi1}) between $\delta$ and $\T$ and requiring  it to be invariant in the limit $L\to\infty$, we obtain the same condition as eq.~(\ref{eqGfin}) with 
$\G_n^{tot}$ replaced by
$K_n$.

\section{Discussion} 
\label{sec:disc} 

We have proposed a new perturbative approach to 
non-linear structure formation 
that overcomes some of the drawbacks 
of standard Eulerian perturbation theory.
Rather than studying the time evolution of cosmological fields, we consider the time evolution of their probability distribution, 
as is common in statistical mechanics.
Starting from generic hydrodynamical equations of motions 
for the density and velocity fields 
we derive the analogue of the Liouville continuity equation that governs 
the time evolution of the probability distribution function.
We have shown that this equation can formally be solved exactly, without any need
for assuming an ad-hoc truncation.
Equal-time observables such as the power spectrum can be computed perturbatively
based on the time-evolved probability distribution.
At this stage time plays the role of an external parameter, which suggests the name Time-Sliced Perturbation
Theory (TSPT).
We have developed the diagrammatic technique within this approach and worked out its ingredients for
the Zel'dovich approximation and the exact Eulerian dynamics
with an initially Gaussian distribution. 
The formulation proposed here admits a representation in terms of a
generating functional for equal-time correlation functions, that is formally similar
to QFT in three-dimensional Euclidean space, and therefore allows to apply QFT techniques
in a transparent way.

An important property of this formulation is that all building blocks entering the
perturbative computation are by themselves infrared safe. This is a direct consequence
of the fact that time-evolution and statistical averaging are disentangled. The perturbative
expansion then involves only equal-time quantities that are protected by the underlying symmetries
related to the equivalence principle.
Compared to the standard Eulerian formulation of perturbation theory (SPT), this facilitates
the application of diagrammatic resummation techniques.

In this work we focussed on the general formalism, demonstrated the infrared safety, and worked out a sample calculation for an ideal pressureless fluid.
However, we paid special attention to keep the derivation generic. It can indeed be easily adapted to describe non-ideal fluids and is particularly suited to include primordial
non-Gaussianity. Furthermore, if the hydrodynamical evolution equations were supplemented
by a stochastic force (noise term), this could be accounted for by replacing the Liouville
equation for the distribution function by a Fokker-Planck equation.

The properties discussed above suggest that TSPT is suitable to address shortcomings of
perturbation theory, in particular, infrared resummation, which is important to capture the
shape of the baryon acoustic peak \cite{BAOnext}. 
This resummation can be systematically formulated at the level
of Feynman diagrams within TSPT, by virtue of its advantageous infrared properties. Furthermore,
due to the similarities with Euclidean QFT, 
it is a natural framework to apply methods of renormalization
group evolution in order to better understand the UV sensitivity of the perturbative expansion.

For realistic applications the method should be extended
to include redshift space distortions and the effects of baryons.
We leave this for future research.
Let us note that the formulation
based on a time-evolved probability distribution function appears 
convenient for the description of biased tracers.

\section*{Acknowledgements}

We thank D.~Gorbunov, E.~Pajer, R.~Porto, P.~De Los Rios, F.~Vernizzi and
Y.~Zenkevich for useful discussions. 
D.B. acknowledges the hospitality of CCPP at NYU while this work was
in preparation. S.S. thanks Kavli Institute for Theoretical Physics at Santa
Barbara for hospitality during the initial stage of this project (the
visit to KITP was supported in part by the National Science Foundation
under Grant NSF PHY11-25915).  
This work was supported by the Swiss National Science Foundation
(M.I. and S.S.)
and the RFBR grants 14-02-31435 and 14-02-00894 (M.I.).

\appendix 
 
\section{TSPT kernels from SPT}
\label{app:spttotspt}

Here we show how the quantities appearing in TSPT can be related to the SPT kernels $F_n$, $G_n$. 
In SPT one expresses the fields at a given moment of time in terms of their initial values, 
\bseq\label{eq:spt}
\begin{align}
&\delta_\e(\k)=\sum_{n=1}e^{n\eta}\int [dq]^n F_n(\q_1,...,\q_n)\,\delta^{(3)}\Big(\k-\sum_{i=1}^n \q_i\Big)\prod_{j=1}^n\T_0(\q_j),\\
&\T_\e(\k)=\sum_{n=1}e^{n\eta}\int [dq]^n G_n(\q_1,...,\q_n)\,\delta^{(3)}\Big(\k-\sum_{i=1}^n  \q_i\Big)\prod_{j=1}^n\T_0(\q_j).
\label{eq:Gn}
\end{align}
\eseq
To connect the $K_n$ kernels with the $F_n$ and $G_n$ kernels consider eq.~\eqref{eq:psi1}.
This produces a hierarchy of equations for the $K_n$ kernels, that can be solved iteratively. For instance, the first non trivial relations are,
\bseq
\label{eq:FGtoK}
\begin{align}
&K_2(\q_1,\q_2)=2\big(F_2(\q_1,\q_2)-G_2(\q_1,\q_2)\big),\\
&K_3(\q_1,\q_2,\q_3) = 6\big(F_3(\q_1,\q_2,\q_3)\!-\!G_3(\q_1,\q_2,\q_3)\big)
-2\!\! \!\!\sum_{1\leq i<j\leq 3} \!\!G_2(\q_i,\q_j)\, K_2\big(\q_i+\q_j,\q_l)\bigg|_{l\neq i,j}
\end{align}
\eseq
The formulas for $I_n$ are derived by substituting the expression \eqref{eq:Gn} into the equations \eqref{eq:eomp} and \eqref{eq:ik}. For the lowest kernels one has,
\bseq
\label{eq:GtoI}
\begin{align}
&I_2(\q_1,\q_2)=2G_2(\q_1,\q_2),\\
&I_3(\q_1,\q_2,\q_3) = 12 G_3(\q_1,\q_2,\q_3)-
4\sum_{1\leq i<j\leq 3}
G_2(\q_i,\q_j)
G_2(\q_i+\q_j,\q_l)\bigg|_{l\neq i,j} \,.
\end{align}
\eseq
Since the vertices $\bar \G_n$ and counterterms $C_n$ are expressed in TSPT through $I_n$ and $K_n$ the above relations allow to write them as combinations of the SPT kernels.

\section{1-loop results and comparison with SPT}\label{app:1loop}

In this appendix we perform an explicit computation of 1-loop corrections to the $\T$ and $\delta$ power spectra in TSPT.
First we summarize the expressions for the vertices that we will need in the calculation. Using the general formula (\ref{solGaus2}) we obtain the 3- and 4-point vertices, 
\bseq
\label{eq:Wvert}
\begin{align}
\label{Wvert3}
&\bar\G_3(\k_1,\k_2,\k_3)=-\delta^{(3)}\Big(\sum \k_i\Big)
\left[\frac{I_2(\k_1,\k_2)}{\bar P(|\k_3|)}+\text{cycl.}\right]\,,\\
&\bar\G_4(\k_1,\k_2,\k_3,\k_4)
=\delta^{(3)}\Big(\sum \k_i\Big)
\bigg\{\frac{1}{2}\sum_{1\leq i<j\leq 4}
I_2(\k_i,\k_j)\bigg[\frac{I_2(\k_i+\k_j,\k_l)}{\bar P(|\k_m|)}
+\text{cycl.}\bigg]\bigg|_{\begin{smallmatrix}
l<m\\
l,m\neq i,j
\end{smallmatrix}
}\notag\\
&\qquad\qquad\qquad\qquad\qquad\qquad\qquad~~~
-\frac{1}{2}\bigg[\frac{I_3(\k_1,\k_2,\k_3)}{\bar P(|\k_4|)}
+\text{cycl.}
\bigg]\bigg\}\,,
\end{align}
\eseq
where ``cycl.'' stands for contributions differing from the first term
in the brackets by cyclic permutations of the momenta. 
For the counterterms we get from (\ref{counter}),
\begin{subequations}
\label{eq:C12}
\begin{align}
\label{eq:C1}
&C_1(\k)= \delta^{(3)}(\k)\int [dq]I_2(\q,\k)\,,\\
&C_2(\k_1,\k_2)= \frac{1}{2}
\delta^{(3)}(\k_1+\k_2)\int
[dq]\Big(I_3(\q,\k_1,\k_2)-I_2(\q,\k_1+\k_2)I_2(\k_1,\k_2)\Big)\,.
\label{eq:C2} 
\end{align}
\end{subequations}
The last term in $C_2$ vanishes if the kernel $I_2$ satisfies
\be
\label{eq:cond_1}
 \lim_{p\to 0}I_2(\q,-\q+{\bf p})I_2(\k,{\bf p})=0 \,,
\ee
which is indeed the case both for ZA and ED.

The 1-loop corrections to the $\T$ power spectrum are given by the diagrams shown in (\ref{diagr1loop0}). The first --- {\it daisy} --- contribution reads,
\be
\label{eq:1-loopsuned}
\begin{split}
&P_{daisy}(\eta;|\k|)=-e^{4\eta} \int [dq] \Bigg[\left(I_2(-\k,\k-\q)I_2(-\k,\q)-\frac{1}{2}I_3(\k,-\k,\q)\right)\bar P(|\k|)^2\\
&+\left(I_2(-\q,-\k+\q)I_2(\k,-\q)-\frac{1}{2}I_3(\k,\q,-\q)\right)\bar P(|\q|)
\bar P(|\k|)+I_2^2(\k,-\q)\frac{\bar P(|\q|)\bar P(|\k|)^2}{\bar P(|\k-\q|)} \Bigg]\,.
\end{split}
\ee
The second term, or {\it fish} diagram, gives,
\be
\label{eq:1-loopfished}
\begin{split}
P_{fish}(\eta;|\k|)=e^{4\eta} 
\int [dq] \Bigg[&I_2(\k,-\q)I_2(\k,\q-\k)\bar P(|\k|)^2
+\frac{I_2(\q,\k-\q)^2}{2}\bar P(|\q|)\bar P(|\q-\k|)\\ 
+I_2(\k,-\q)I_2(\k-&\q,\q)\bar P(|\q|)\bar P(|\k|)
+I_2(\q-\k,\k)I_2(\k-\q,\q)\bar P(|\k-\q|)\bar P(|\k|)\\
&+\frac{I_2^2(\q-\k,\k)}{2}\frac{\bar P(|\q-\k|)\bar P(|\k|)^2}{\bar P(|\q|)}
+\frac{I_2^2(\k,-\q)}{2}\frac{\bar P(|\q|)\bar P(|\k|)^2}{\bar P(|\k-\q|)}
\Bigg]\,.
\end{split}
\ee
Notice the presence of terms with the linear power spectrum $\bar P(|\k|)$ in the denominator. They arise as a result of the $1/\bar P$ dependence of the TSPT vertices. Such terms cannot appear in SPT, which is an analytic expansion in  $\bar P(|\k|)$. Indeed, one verifies explicitly, that these terms cancel in the sum of   
(\ref{eq:1-loopsuned}) and (\ref{eq:1-loopfished}). Another difference from SPT are  the first terms in (\ref{eq:1-loopsuned}), (\ref{eq:1-loopfished}) proportional to $\bar P(|\k|)^2$. They contain UV divergent momentum integrals which are independent of the power spectrum. The contributions with the kernels $I_2$ cancel between (\ref{eq:1-loopsuned}) and (\ref{eq:1-loopfished}), whereas the contribution with $I_3$ is canceled by the third --- {\it counterterm} --- diagram from (\ref{diagr1loop0}),
\be
P_{counter}=-\frac{e^{4\eta}}{2}\bar P(|\k|)^2\int [dq]I_3(\k,-\k,\q)\,.
\ee
The rest of terms are brought to the form,
\be
\label{eq:1-looptotal}
\begin{split}
P^{\rm 1-loop}_{\T\T}(\eta;|\k|)&=e^{4\eta}  \int [dq]\Bigg[
\frac{1}{2}\bar P(|\k|)\bar P(|\q|)I_3(\k,\q,-\q)\\
+ \bar P(|\k|)&\bar P(|\k-\q|)I_2(\q-\k,\k)I_2(\k-\q,\q)
+\frac{I_2^2(\q,\k-\q)}{2}\bar P(|\q|)\bar P(|\k-\q|)\Big] \,.
\end{split}
\ee
Making use of eqs.~(\ref{eq:FGtoK}), (\ref{eq:GtoI}) one recovers 
the SPT result,
\be
\begin{split}
\label{eq:PSthZA}
P_{\T\T}^{\rm1-loop}&(\eta;|\k|)=
2\int [dq]
\Big[3G_3(\k,-\q,\q)P_\e(|\q|)P_\eta(|\k|)+G_2^2(\k-\q,\q)P_\e(|\q|)P_\eta(|\k-\q|)
\Big],
\end{split}
\ee
where $P_\e(|\k|)$ is the linear power spectrum at the time $\e$.

To find the correction to the density power spectrum we need to add the three `composite operator' diagrams (\ref{eq:Ddelta}). 
They read,
\bseq
\begin{align}
&\hat P_{(i)}(\e;|\k|)\!\!=\!e^{4\e}\!\!\!\int [dq]\Big[2I_2(\k,-\q)\bar P(|\k|)\bar P(|\q|)
+I_2(\q,\k-\q)\bar P(|\q|)\bar P(|\k-\q|)\Big] K_2(\q,\k-\q)\,,\\
&\hat P_{(ii)}(\e;|\k|)=e^{4\e}\int [dq]\frac{K_2(\q,\k-\q))^2\bar P(|\k-\q|)\bar P(|\q|)}{2}\,,\\
&\hat P_{(iii)}(\e;|\k|)=e^{4\e}\int [dq]K_3(\q,-\q,\k)\bar P(|\q|)\bar P(|\k|)\,.
\end{align}
\eseq
Adding these contributions to $P_{\T\T}^{\rm 1-loop}$ and using 
(\ref{eq:FGtoK}), (\ref{eq:GtoI}) we arrive at the SPT expression,
\be
\begin{split}
P^{\rm 1-loop}_{\delta\delta}(\e;|\k|)=&6P_\e(|\k|)\int [dq] F_3(\k,\q,-\q)P_\e(|\q|)\\
&+2\int [dq] \left[F_2(\q,\k-\q)\right]^2P_\e(|\q|)P_\e(|\k-\q|)\,.
\end{split}
\ee

\section{IR safety and the initial power spectrum}
\label{app:initialPS}

The TSPT vertices $\bar \G_n$ contain terms that have the initial power spectrum in the denominator and may, in principle, have IR divergence depending on its slope. 
Recall that these terms have to cancel in the final expressions for correlation functions. 
Thus, the IR singularities of $\G_n$ related to the slope of the power spectrum are spurious and cannot affect physical observables.
We would like to mention that even in the presence of these 
singularities the statement of Sec.~\ref{sec:IRsafe} that individual 
loop diagrams in TSPT are IR convergent if $\bar P(|\q|)\propto q^\nu$ with
$\nu > -3$ at $q\to 0$ holds true, because $\G_n$ - vertices have to be multiplied by 
loop propagators inside loops so that eventually the singular contributions are
completely cancelled.

The 2-point vertex $\bar \G'_2(-\q,\q)$ is finite provided that the initial power spectrum 
does not vanish anywhere. In the real universe, however, $\bar P(|\q|)\propto q$ at $q\to 0$ and
thus, formally $\bar \G'_2\sim 1/q$ in the soft limit. We will now show that the divergence related to the
$1/P$ terms are absent in higher order vertices if at least two of
their arguments are hard\footnote{A configuration with a single hard
  momentum is forbidden by the momentum conservation.} and the power
spectrum behaves as $\bar P(|\q|)\propto q^\nu$ with $\nu\leq 2$ at
$q\to 0$. This includes the case of the real universe where $\nu=1$.

Let us start with the $\bar \G_3$ vertex in the limit where two
momenta, $\k$, $(-\k-\q)$,
are hard and the third one is soft, $\q \equiv \epsilon \q'$, $\epsilon \to 0$. 
We use the expression \eqref{identificED2} and focus on the term $\sim 1/\bar P(|\q|)$
(the IR divergences cancel for other terms, see Sec. \ref{sec:IRsafe}),
\be
\bar \G'_3(-\k-\q,\k,\q)\ni \frac{I_2(-\k-\q,\k)}{\bar P(|\q|)} \propto \frac{q^2}{k^2}\frac{1}{\bar P(|\q|)}\,.
\ee
We see that there is no IR divergence 
provided that $\bar P(|\q|)\propto q^{\nu}$ with $\nu\leq 2$ at $q\to 0$.
This result can be readily generalized to an arbitrary vertex 
with a number of soft and 
hard momenta. 
Let us prove it by induction in the case of two hard momenta and all
other momenta uniformly sent to zero 
as in \eqref{softlimit}. As an induction hypothesis we assume that
the terms $\sim 1/\bar P(|\q_i|)$ entering the vertex 
$\bar \G'_{n-1}(\k,-\k-\sum_{i=1}^{n-3} \q_i,\q_1,...,\q_{n-3})$ 
scale as $\epsilon^2/P(\epsilon|\q'|)\sim \epsilon^2/P(\epsilon)$. 
We will use the shorthand $\q\equiv \sum_{i=1}^{n-2} \q_i$.
It is convenient to divide the vertex $\G'_{n}$ into two pieces,
\be
\bar \G'_n=-\frac{1}{n-2}\left(\bar \G'_{n,A} +\bar \G'_{n,B}\right)\,,
\ee
where $\bar \G'_{n,A}$ is the piece of the recursion relations \eqref{Gned},\eqref{recurG} common in ZA and ED, and $\bar \G'_{n,B}$
is a contribution appearing in ED (second line of \eqref{Gned}). 
One has, 
\begin{subequations}
\begin{align}
\label{eqind:line1}
\bar \G'_{n,A}(\k,-\k-\q,\q_1,...,\q_{n-2})=&
\sum_{j=1}^{n-2} I_2(\q_j,\k)\bar \G'_{n-1}(\k+\q_j,...,\check{\q}_j,...)\\
\label{eqind:line2}
+&\sum_{j=1}^{n-2} I_2(\q_j,-\k-\q)\bar \G'_{n-1}(\k,-\k-\q+\q_j,...,\check{\q}_j,...)\\
\label{eqind:line3}
+&\sum_{l>j=1}^{n-2} I_2(\q_j,\q_l)
\bar \G'_{n-1}(...,\q_j+\q_l,...,\check{\q}_l,...)\\
\label{eqind:line4}
+&I_2(\k,-\k-\q)
\bar \G'_{n-1}(-\q,\q_1,...,\q_{n-2})\,.
\end{align}
\end{subequations}
As was proven in Sec. \ref{sec:IRsafe}, the IR divergences appearing in the first two terms 
\eqref{eqind:line1},\eqref{eqind:line2} cancel up to $O(\epsilon^0)$ order. According to the induction hypothesis
the corresponding vertices already contain the terms $\sim \epsilon^2/\bar P(\epsilon)$ and thus satisfy the hypothesis. 
The term \eqref{eqind:line3} diverges only as $\sim \epsilon^2/\bar P(\epsilon)$
because $I_2(\q_i,\q_j)\sim O(\epsilon^0)$. 
Finally, the term \eqref{eqind:line4} features the vertex scaling as $1/\bar P(\epsilon )$ 
that multiplies the following $I_2$ kernel,
\be
 I_2(\k,-\k- \q)= O\left({q^2}/{k^2}\right)=O(\epsilon^2)\,.
\ee
Thus, we have checked that the $\bar \G_{n,A}$ piece satisfies the induction hypothesis.
Now we turn to the case of $\bar \G_{n,B}$. 
This contribution contains the $K_n$ kernels which were proven to be at least $O(\epsilon^0)$
for any momenta configuration (see \eqref{eq:KIRs}). Thus, if one of those multiplies the $\bar \G'_n(\k,-\k,...)$ vertex, it will not 
produce any IR enhancement. 
The only term that does not obviously satisfy the induction hypothesis
is 
\begin{subequations}
\begin{align}
\nonumber
&\bar \G_{n,B}(\k,-\k-\q,\q_1,...,\q_{n-2})\ni \\
\nonumber
&\frac{3}{2}\sum_{m=3}^{n-1}\frac{1}{(m-2)!(n-m)!}\sum_\sigma 
K_m(\k,-\k-\q,\q_{\sigma(1)},...,\q_{\sigma(m-2)}) \\
\label{eqind:line6}
&\quad \quad\quad\times \bar \G'_{n-m+1}(-\sum_{l=m-2}^{n-2}\q_{\sigma(l)},\q_{\sigma(m-2)},...,\q_{\sigma(n-2)})
\end{align}
\end{subequations}
In order to show that this term scales as $\epsilon^2/P(\epsilon)$
it is sufficient to prove that the $K_m$ kernels go as $\sim \epsilon^2$ in the limit where two 
hard momenta almost satisfy momentum conservation,
\be
\label{indind}
\lim_{\epsilon \to 0} K_m(\k,-\k-\epsilon\q',\epsilon \q'_1,...,\epsilon \q'_{m-2})=O(\epsilon^2)\,, \; m \geq 3\,.
\ee
This behavior is obviously satisfied for $m=2$, see (\ref{eq:K2ED}). 
Suppose now the above is true for some $m-1$. 
Recall that in Sec.~\ref{sec:5.1} we have
proven that the kernels $K_n$ are bounded for {\em all} values of
momenta. Then using \eqref{Knsol} one can readily see that most of the terms in the
recursion relation satisfy the induction hypothesis because the corresponding kernels have the same momenta 
configuration as \eqref{indind} and multiply some other
$O(\epsilon^0)$ terms. 
The rest is given by
\begin{subequations}
\begin{align}
\label{line1}
&\frac{(2n+3)}{2}K_m(\k,-\k-\q,\q_1,...,\q_{m-2})\ni \nonumber\\
&\alpha(\k,-\k-\q+\tilde\q)K_{m-1}(-\k-\q,\q_1,...)+
\alpha(-\k-\q,\k+\tilde\q)K_{m-1}(\k,\q_1,...)\\
\label{line2}
&-\sum_{i=1}^{m-2} I_2(\k,\q_i)K_{m-1}(\k+\q_i,-\k-\q,\check{\q}_j)
-\sum_{i=1}^{m-2} I_2(-\k,\q_i)K_{m-1}(-\k-\q+\q_i,\k,\check{\q}_j)\,,
\end{align}
\end{subequations}
where $\tilde\q\equiv \sum_{i=1}^{m-2}\q_i$.
One can show that the sum of the terms \eqref{line1} is $O(\epsilon^2)$
as a consequence of\footnote{This property can be straightforwardly proven by induction. Indeed, if one assumes Eq.\eqref{eq:Kmarg} is true for all $l\leq m-1$, one can write down the recursion relation \eqref{Knsol} that generates $K_m(\k,\q_1,...)$. 
Then one can notice that after the cancellation of IR dangerous contributions (see Eq.\eqref{eq:KIRs})
the remainders of the corresponding $\alpha$ and $\beta$ kernels are even functions of $\k$ at order $O(\epsilon^0)$, and thus,
to this order,
one can safely substitute all $K_{l}(\k,\q_1,...)$($l\leq m-1$) with $K_{l}(-\k,\q_1,...)$ in the recursion relation, which will immediately yield \eqref{eq:Kmarg}.} 
\be
\label{eq:Kmarg}
K_m(\k,\q_1,...,\q_{m-1})= K_m(-\k,\q_1,...,\q_{m-1})+O(\epsilon).
\ee
Finally, in 
\eqref{line2} the pole contributions from the $I_2$ kernel cancel up to 
$O(\epsilon^0)$ so that the residual term remains $O(\epsilon^2)$ because of the induction hypothesis.
This completes the proof for the case of two hard momenta. 
Extension to the case of more hard momenta is straightforward.

In summary, we have shown that the $\bar \G'_n$ vertices with at least
some of their arguments hard
do not contain any divergences in the IR limit provided that 
$\bar P(k)\propto k^{\nu},\;\nu\leq 2$ at $k\to 0$, which is the case for realistic cosmology.


\end{document}